\documentclass{elsart}
\usepackage{epsfig,latexsym,amsfonts}
\newcommand{\cO}{{\mathcal O}}
\newcommand{\third}{\mbox{\small $\frac{1}{3}$}}
\newcommand{\Dd}[1]{\mbox{
  \parbox[b]{0cm}{$D$}\raisebox{1.7ex}{$\leftrightarrow$}$_{\!#1}$}}
\makeatletter
\@addtoreset{equation}{section}
\makeatother

\begin{document}
\begin{frontmatter}
 
\title{
\vspace{-1.9cm}
\hfill {\normalsize \mdseries DESY 98-097} \\[-0.2cm] 
\hfill {\normalsize \mdseries TPR-98-19} \\[-0.2cm] 
\hfill {\normalsize \mdseries HUB-EP-98/45} \\[0.4cm] 
Nonperturbative Renormalisation of Composite Operators in  
           Lattice QCD}

\author[Regensbg]{M. G\"ockeler},
\author[HU]{R. Horsley},
\author[Zeuthen]{H. Oelrich},
\author[Leipzig]{H. Perlt},
\author[Zeuthen,FU]{D. Petters},
\author[Regensbg]{P.E.L. Rakow},
\author[Regensbg]{A. Sch\"afer},
\author[Zeuthen,Hamburg]{G. Schierholz},
\author[Leipzig]{A. Schiller}
\address[Regensbg]{Institut f\"ur Theoretische Physik,
         Universit\"at Regensburg, D-93040 Regensburg, Germany}
\address[HU]{Institut f\"ur Physik, Humboldt-Universit\"at,
         D-10115 Berlin, Germany}
\address[Zeuthen]{Deutsches Elektronen-Synchrotron DESY, 
         D-15735 Zeuthen, Germany}
\address[Leipzig]{Institut f\"ur Theoretische Physik,
         Universit\"at Leipzig, D-04109 Leipzig, Germany}
\address[FU]{Institut f\"ur Theoretische Physik,
             Freie Universit\"at Berlin, D-14195 Berlin, Germany}
\address[Hamburg]{Deutsches Elektronen-Synchrotron DESY, 
                  D-22603 Hamburg, Germany}

\begin{abstract}
We investigate the nonperturbative renormalisation of composite
operators in lattice QCD restricting ourselves to operators
that are bilinear in the quark fields.
These include operators which are relevant to the calculation of 
moments of hadronic structure functions. The computations are based on
Monte Carlo simulations using quenched Wilson fermions.
\end{abstract}
\begin{keyword}
Lattice QCD; nonperturbative renormalisation; composite operators;
operator product expansion
\PACS{11.10.Gh; 11.15.Ha; 12.38.Gc}
\end{keyword}

\end{frontmatter}
 
\section{Introduction}
 
Monte Carlo simulations of lattice QCD have evolved from spectrum
calculations to more detailed investigations of hadron
structure. These more advanced studies require to calculate
hadronic matrix elements of composite operators. In order to compute
the moments of hadronic structure functions, for example, one needs
the matrix elements of composite operators appearing in the 
operator product expansion of the appropriate currents 
\cite{mart,capi,roma,letter}. In general, one has to renormalise 
these operators in order to obtain finite answers in the continuum
limit. Furthermore, comparison with the results of phenomenological
analyses usually requires the matrix elements to be given in one of
the popular continuum renormalisation schemes, e.g.\ the 
$\overline{\mathrm{MS}}$ scheme. So one has to think about the
conversion of the bare lattice operators to renormalised continuum
operators. 

Consider the special case of the operators determining the
moments of ha\-dron\-ic structure functions. Here the renormalised 
continuum matrix element has to be multiplied by the corresponding 
Wilson coefficient to yield the desired moment, which can be measured
in deep inelastic scattering experiments.
Being an observable quantity it must be independent of
any choices made in the renormalisation procedure. This
is of course only possible, if the renormalised matrix element and 
the Wilson coefficient are calculated in the same scheme. 
In particular, the
dependence on the renormalisation scale $\mu$ has to cancel between
the renormalisation constant and the Wilson coefficient. 
Since the Wilson coefficients are usually computed in the 
$\overline{\mathrm{MS}}$ scheme, we have to convert our lattice
operators to continuum $\overline{\mathrm{MS}}$ operators if we want to
make contact with phenomenology.

One obvious possibility to calculate the necessary renormalisation
factors is lattice perturbation theory. However, quite often lattice
perturbation theory seems to converge rather slowly. Identifying one
source of these poor convergence properties, Lepage and Mackenzie
proposed as a remedy the so-called tadpole improved perturbation
theory \cite{lepmac}. Still, lattice perturbation series rarely 
extend beyond the one-loop level, and hence considerable 
uncertainty remains. 

It is therefore only natural to try a nonperturbative renormalisation
by means of Monte Carlo simulations. A way how to do this was
introduced in Ref.~\cite{marti}. We shall present
several improvements of this method and apply it to a variety
of operators that are bilinear in the quark fields~\cite{thesis}. 
In particular, we study operators which are needed to calculate 
hadronic structure functions. Computing the $Z$ factors for a 
rather large range of renormalisation scales we 
can find out at which scales (if at all) perturbative behaviour 
sets in such that the multiplication with the perturbative Wilson
coefficients makes sense. 

In this paper we shall work only with Wilson fermions in the quenched 
approximation. However, an obvious next step is the use of
improved fermions in order to reduce cut-off effects. Like the
renormalisation of our operators, the improvement should be
nonperturbative. As far as the action is concerned, there is already
a lot of experience how to achieve this. But improvement of
the action is not sufficient, the operators have to be improved
as well. Here a good deal of work still has to be done, in particular
for operators containing derivatives.

The paper is organised as follows: After introducing in 
Section~\ref{sec.operators} the operators to be studied we explain the
method of nonperturbative renormalisation in Section~\ref{sec.method}.
In the special case of the vector and axial vector currents we 
prefer a variant of this method because it seems to suppress 
lattice artifacts more efficiently. We describe it in 
Section~\ref{sec.vector}. 
Perturbative formulae, which we need for comparison
and for obtaining results in the $\overline{\mathrm{MS}}$ scheme, are
given in Section~\ref{sec.pert}. Section~\ref{sec.numerics} contains
some details of the numerical implementation including our momentum
sources which greatly reduce the statistical noise of the data.
In Section~\ref{sec.results}
we present and discuss our results. Finally, we come to our conclusions.

\section{The operators} \label{sec.operators}

In the Euclidean continuum we should study the operators 
\begin{equation}
  \cO ^{(q)}_{\mu_1 \cdots \mu_n} =   
        \bar{q}\gamma_{\mu_1} \Dd{\mu_2} \cdots \Dd{\mu_n} q \,, 
\end{equation}
\begin{equation}
  \cO ^{5(q)}_{\sigma \mu_1 \cdots \mu_n} =   
        \bar{q}\gamma_\sigma \gamma_5 \Dd{\mu_1} \cdots \Dd{\mu_n} q 
\end{equation}
or rather O(4) irreducible multiplets with definite C-parity. In
particular, we obtain twist-2 operators by symmetrising the indices
and subtracting the traces. In the flavour-nonsinglet case they do not
mix and are hence multiplicatively renormalisable. 

Working with Wilson fermions it is straightforward to write
down lattice versions of the above operators. One simply replaces the
continuum covariant derivative by its lattice analogue. However, O(4)
being restricted to its finite subgroup H(4) (the hypercubic group) on the
lattice, the constraints imposed by space-time symmetry are less
stringent than in the continuum and the possibilities for mixing increase
\cite{capi,roma,grouptheory,pertz}. Guided by the H(4) classification 
given in Ref.~\cite{grouptheory} we have chosen the operators
\cite{letter,pirho,a1paper}
\begin{eqnarray}
\cO_{v_{2,a}}  & = &   \cO ^{(q)}_{\{14\}} \,,  \\
\cO_{v_{2,b}}  & = &   \cO ^{(q)}_{\{44\}} - \third ( \cO ^{(q)}_{\{11\}}
                    + \cO ^{(q)}_{\{22\}} + \cO ^{(q)}_{\{33\}} ) \,,   \\    
\cO_{v_3}      & = &   \cO ^{(q)}_{\{114\}} - \half ( \cO ^{(q)}_{\{224\}}
                    + \cO ^{(q)}_{\{334\}} )  \,,    \\    
\cO_{v_4}      & = &   \cO ^{(q)}_{\{1144\}} + \cO ^{(q)}_{\{2233\}}
                    - \cO ^{(q)}_{\{1133\}} - \cO ^{(q)}_{\{2244\}} \,, \\    
\cO_{a_1}      & = &   \cO ^{5(q)}_{\{24\}} \,,   \\
\cO_{a_2}      & = &   \cO ^{5(q)}_{\{214\}} \,,   \\
\cO_{r_{2,a}}  & = &   \cO ^{5(q)}_{\{14\}} \,,  \\
\cO_{r_{2,b}}  & = &   \cO ^{5(q)}_{\{44\}} - \third ( \cO ^{5(q)}_{\{11\}}
                    + \cO ^{5(q)}_{\{22\}} + \cO ^{5(q)}_{\{33\}} ) \,,   \\ 
\cO_{r_3}      & = &   \cO ^{5(q)}_{\{114\}} - \half ( \cO ^{5(q)}_{\{224\}}
                    + \cO ^{5(q)}_{\{334\}} )  \,.
\end{eqnarray}
They are labeled by the reduced hadron matrix elements which they determine. 
In addition, we have studied the following operators without derivatives
(``currents''):
\begin{eqnarray}
\cO^S  & = &   \bar{q} q \,, \\
\cO^P  & = &   \bar{q} \gamma_5 q \,, \\
\cO^V_\mu  & = &   \bar{q} \gamma_\mu q \,, \\
\cO^A_\mu  & = &   \bar{q} \gamma_\mu \gamma_5 q \,,
\end{eqnarray}
where all quark fields are taken at the same lattice point, 
as well as the conserved vector current
\begin{eqnarray}
 J_\mu (x) & = &  \half \Big( \bar{q} (x+\hat{\mu}) (\gamma_\mu+1) 
  U^+ (x,\mu) q(x)  \nonumber \\ 
 {} & {} & \hspace{3.0cm} 
     {} + \bar{q} (x) (\gamma_\mu-1) U(x,\mu) q(x+\hat{\mu})
  \Big) 
\end{eqnarray}
with the link matrix $U(x,\mu) \in$ SU(3) representing the gauge field.

Note that the operators $\cO_{v_{2,a}}$ and $\cO_{v_{2,b}}$     
although belonging to the same irreducible O(4) multiplet transform
according to inequivalent representations of H(4). Hence their
renormalisation factors calculated on the lattice have no reason to
coincide. The same remark applies to $\cO_{r_{2,a}}$ and $\cO_{r_{2,b}}$.
The operators $\cO_{a_1}$ and $\cO_{r_{2,a}}$, on the other hand,
are members of the same irreducible H(4) multiplet, hence their
renormalisation factors should agree also on the lattice.

Concerning the mixing properties a few remarks are in order. 
Mixing with operators of equal or lower dimension is excluded for the
operators $\cO_{v_{2,a}}$, $\cO_{v_{2,b}}$, 
$\cO_{a_1}$, $\cO_{a_2}$, $\cO_{r_{2,a}}$, $\cO_{r_{2,b}}$, 
as well as for the currents.
The case of the operator $\cO_{v_3}$, 
for which there are two further operators with the same
dimension and the same transformation behaviour, is discussed in 
Refs.~\cite{grouptheory,pertz}. The operators $\cO_{v_4}$, $\cO_{r_3}$,
on the other hand, could in
principle mix not only with operators of the same dimension but also
with an operator of one dimension less and different chiral properties.
It is of the type
\begin{equation}
 \bar{q} \sigma_{\mu \nu} \gamma_5
 \Dd{\mu_1} \Dd{\mu_2} \cdots \Dd{\mu_n} q \,,
\end{equation}
where $n=2$ in the case of $\cO_{v_4}$
and $n=1$ for $\cO_{r_3}$. 

Our analysis ignores mixing completely. This seems to be 
justified for $\cO_{v_3}$. Here a perturbative calculation gives a rather
small mixing coefficient for one of the mixing operators \cite{roma,pertz}, 
whereas the
other candidate for mixing does not appear at all in a one-loop
calculation of quark matrix elements
at momentum transfer zero, because its Born term 
vanishes in forward direction. 
The same is true for all operators of dimension less or equal to 6
which transform identically to $\cO_{v_4}$: Their Born term vanishes
in forward matrix elements, hence they do not show up in a one-loop
calculation at vanishing momentum transfer. 
In the case of $\cO_{r_3}$, however, the mixing with an
operator of lower dimension is already visible at the one-loop level
even in forward direction.

\section{The method} \label{sec.method}

Our calculation of renormalisation constants follows closely the
procedure proposed by Martinelli et al.\ \cite{marti}. It mimics the
definitions used in (continuum) perturbation theory. We work
on a lattice of spacing $a$ and volume $V$ in Euclidean space.
For a fixed gauge let
\begin{equation} \label{Gdef}
 G_{\alpha\beta} (p) = \frac{a^{12}}{V} \sum_{x,y,z} 
      {\mathrm e}^{- {\mathrm i} p \cdot (x-y) } 
      \langle q_\alpha (x) \cO (z) \bar{q}_\beta (y) \rangle
\end{equation}
denote the non-amputated quark-quark Green function with one insertion
of the operator $\cO$ at momentum zero. It is to be considered as a
matrix in colour and Dirac space. With the quark propagator
\begin{equation}
 S_{\alpha\beta} (p) = \frac{a^8}{V} \sum_{x,y} {\mathrm e}^{- {\mathrm i} 
      p \cdot (x-y) } \langle q_\alpha (x) \bar{q}_\beta (y) \rangle 
\end{equation}
the corresponding vertex function (or amputated Green function) is
given by 
\begin{equation}
\Gamma (p) = S^{-1} (p) G(p) S^{-1} (p) \,. 
\end{equation}
Defining the renormalised vertex function by 
\begin{equation}
\Gamma_{\mathrm R} (p) = Z_q^{-1} Z_\cO \Gamma (p)
\end{equation}
we fix the renormalisation constant $ Z_\cO $ by imposing the
renormalisation condition
\begin{equation} \label{defz}
 \mbox{\small $\frac{1}{12}$} {\rm tr} \left( \Gamma_{\mathrm R} (p)
   \Gamma_{\mathrm {Born}} (p) ^{-1} \right) = 1 
\end{equation}
at $p^2 = \mu^2$, where $\mu$ is the renormalisation scale. 
So we calculate $ Z_\cO $ from
\begin{equation} \label{calcz}
 Z_q^{-1} Z_\cO \mbox{\small $\frac{1}{12}$} {\rm tr} \left( \Gamma (p)
   \Gamma_{\mathrm {Born}} (p) ^{-1} \right) = 1 
\end{equation}
with $p^2 = \mu^2$. Here $\Gamma_{\mathrm {Born}} (p) $ is the 
Born term in the 
vertex function of $\cO$ computed on the lattice, and
$Z_q$ denotes the quark field renormalisation constant. 
The latter can be taken as 
\begin{equation}
 Z_q (p) = \frac{ {\rm tr} \left( - {\rm i} \sum_\lambda \gamma_\lambda 
           \sin (a p_\lambda) a  S^{-1} (p) \right) }
           {12 \sum_\lambda \sin^2 (a p_\lambda) } \,, 
\end{equation}
again at $p^2 = \mu^2$.

If the operator under study belongs to an O(4) multiplet of dimension
greater than 1, i.e.\ if it carries at least one space-time index, the
trace in Eq.~(\ref{calcz}) will in general depend on the direction 
of $p$. So the
renormalisation condition (\ref{defz}) violates O(4) covariance even
in the continuum limit. In the continuum, this disease is easily cured
by a suitable summation over the members of the O(4) multiplet. On the
lattice, this makes sense only in the few special cases where the O(4)
multiplet is irreducible also under the hypercubic group H(4), because
the renormalisation constants for the different H(4) multiplets within
a given O(4) multiplet will in general be different. So we have to live
with this noncovariance, which, of course, should finally be
compensated when we convert our results to a covariant renormalisation
scheme. 

The scale $\mu$ at which our renormalisation constants are 
defined should ideally satisfy the conditions
\begin{equation}
  1/L^2 \ll \Lambda^2_{\mathrm {QCD}} \ll \mu^2 \ll 1/a^2 
\end{equation}
on a lattice with linear extent $L$. Whether in a concrete calculation 
these conditions may be considered as fulfilled remains to
be seen. 

\section{A special case: vector and axial vector currents}
\label{sec.vector}

In the special case of the vector and axial vector currents in 
the continuum we can distinguish between longitudinal and 
transverse components with respect to the momentum.
One may thus define two renormalisation constants, one from
the longitudinal and one from the transverse components. 
Denoting the vertex function of the currents generically by $J_\mu (p)$
we have for the vector current (the modifications required in
the case of the axial vector current are obvious)
\begin{equation} \label{longcon} 
 Z_q^{-1} Z_J ^{\mathrm {long}} \mbox{\small $\frac{1}{12}$}
{\rm tr} \left( \sum_\mu 
               \frac{p_\mu}{p^2} \not\!{p} J_\mu (p) \right) 
                        = 1 \,,
\end{equation}
\begin{equation} \label{transvcon}
 Z_q^{-1} Z_J ^{\mathrm {trans}} \mbox{\small $\frac{1}{12}$}
{\rm tr} \left( \sum_\mu \left( \gamma_\mu -
    \frac{p_\mu}{p^2} \not\!{p} \right) J_\mu (p) \right) 
                        = 3 \,.
\end{equation}
Note that these renormalisation conditions are O(4) covariant.

We try to generalise (\ref{transvcon}) to the lattice
imposing a renormalisation condition of the form
\begin{equation} \label{transv}
 Z_q^{-1} Z_J \mbox{\small $\frac{1}{12}$}
   {\rm tr} \left( \sum_\mu Q_\mu^\bot J_\mu (p) \right)
 = \mbox{\small $\frac{1}{12}$} {\rm tr} \left( \sum_\mu Q_\mu^\bot 
     J_\mu^{\mathrm {Born}} (p) \right) = 3\,,
\end{equation}
where $Q_\mu^\bot$ is a suitable Dirac matrix and 
$J_\mu^{\mathrm {Born}} (p) $ denotes the Born term evaluated on
the lattice. In the case of the
conserved vector current we know that $Z_J = 1$, hence we can 
calculate $Z_q$ from (\ref{transv}). This value is then used 
to determine $Z_J$ for the local vector and axial vector currents.

The matrix $Q_\mu^\bot$ is constructed as 
\begin{equation}
Q_\mu^\bot = \left( J_\mu^{\mathrm {Born}} \right)^{-1} - Q_\mu^\| \,, 
\end{equation}
where
\begin{equation} 
Q_\mu^\| = \frac{1}{N} \sin (a p_\mu) \sum_\lambda 
      \left( J_\lambda^{\mathrm {Born}} \right)^{-1} \sin (a p_\lambda) \,.
\end{equation}
The normalisation factor $N$ follows immediately from the condition
(\ref{transv}):
\begin{equation}
N = \mbox{\small $\frac{1}{12}$} {\mathrm {tr}}
      \left[ \sum_\mu J_\mu^{\mathrm {Born}} \sin (a p_\mu)
      \sum_\lambda \left( J_\lambda^{\mathrm {Born}} \right)^{-1} 
           \sin (a p_\lambda) \right] \,.
\end{equation}
Note that for $a \to 0$ and $J_\mu^{\mathrm {Born}} \to \gamma_\mu$
we have $N/a^2 \to p^2$, and (\ref{transv}) 
reduces to (\ref{transvcon}) in the continuum limit.

In the case of the conserved vector current we find
\begin{equation}
J_\mu^{\mathrm {Born}} (p) = \gamma_\mu \cos (a p_\mu) +
                                \mathrm{i} \sin (a p_\mu)
\end{equation}
so that
\begin{equation}
N = \left( \sum_\mu \sin^2 (a p_\mu) \right)^2 + \sum_\mu \sin^2 (a p_\mu)
    - \sum_\mu \sin^4 (a p_\mu) \,.
\end{equation}
For the local vector and axial vector current we have
\begin{equation}
J_\mu^{\mathrm {Born}} = \gamma_\mu 
\end{equation}
and
\begin{equation}
J_\mu^{\mathrm {Born}} = \gamma_\mu \gamma_5 \,, 
\end{equation}
respectively, and we obtain in both cases 
\begin{equation} \label{normloc}
N = \sum_\mu \sin^2 (a p_\mu) \,.
\end{equation}
Working with the 
longitudinal component, i.e.\ with a lattice version of (\ref{longcon}),
leads to a less smooth momentum 
dependence, i.e.\ to stronger lattice effects.

\section{Input from perturbative calculations} \label{sec.pert}

Eq.~(\ref{defz}) defines a renormalisation scheme
of the momentum subtraction type, which we call MOM scheme.
Note that it will in general not agree with any of the momentum
subtraction schemes used in continuum perturbation theory.
It is desirable to convert our results into a more popular scheme like
the $\overline{\mathrm{MS}}$ scheme. 
Moreover, we want to use our renormalisation factors in connection
with the Wilson coefficients, which appear in the operator product
expansion, and these are generally given
in the $\overline{\mathrm{MS}}$ scheme. Hence we have to perform a
finite renormalisation leading us from the renormalisation scheme
defined by Eq.~(\ref{defz}) to the $\overline{\mathrm{MS}}$ scheme. The
corresponding renormalisation constant 
$Z^{\overline{\mathrm {MS}}}_{\mathrm{MOM}}$
is computed in continuum
perturbation theory using dimensional regularisation.

We work in a general covariant gauge (gauge parameter $\xi$) such
that the gluon propagator has the form
\begin{equation}
  \frac{1}{p^2} \left( \delta_{\mu \nu} - (1 - \xi)
           \frac{p_\mu p_\nu}{p^2} \right) \,.
\end{equation}
The Landau gauge, which is employed in our numerical simulations, 
corresponds to $\xi = 0$. Most of our operators can be
written in the form
\begin{equation}
  \sum_{\mu_1,\ldots,\mu_n} c_{\mu_1 \ldots \mu_n} \left(
  \bar{q}\gamma_{\{\mu_1} \Dd{\mu_2} \cdots \Dd{\mu_n\}} \gamma_5^{n_5} q 
    - \mathrm{traces} \right)
\end{equation}
where $n_5 = 0,1$ and $c_{\mu_1 \ldots \mu_n}$ is 
totally symmetric and traceless. Neglecting quark masses
one obtains with an anticommuting $\gamma_5$
\begin{eqnarray}
Z^{\overline{\mathrm {MS}}}_{\mathrm {MOM}} & = &
    1 + \frac{g^2}{16 \pi^2} C_F \Bigg[ G_n + (1-\xi) S_{n-1}
 \nonumber \\ 
 {} & {} &  {} + \left( - \frac{4}{n+1} + (1-\xi) \frac{2}{n} \right)
      \frac{\left( \sum_\mu p_\mu h_\mu (p) \right)^2 }
      {p^2 \sum_\mu h_\mu (p)^2}    \Bigg] + O(g^4) \,,
\end{eqnarray}
where $C_F=4/3$ for the gauge group SU(3),
\begin{eqnarray} 
  G_n & = & \frac{2}{n(n+1)} \left( -3-S_{n-1}+ 2 S_{n+1} \right) 
  \nonumber \\
  {} & {} & \hspace{3.0cm} {} + \frac{2}{n+1} - 4 \sum_{j=2}^n \frac{1}{j} 
        \left( 2S_j - S_{j-1} \right) -1 \,, \\
  S_n & = & \sum_{j=1}^n \frac{1}{j} \,,
\end{eqnarray}
and
\begin{equation}
 h_\mu (p) = \sum_{\mu_2,\ldots,\mu_n} c_{\mu \mu_2 \ldots \mu_n}
              p_{\mu_2} \cdots p_{\mu_n} \,.
\end{equation}
Because of the noncovariance of our renormalisation condition,
$Z^{\overline{\mathrm{MS}}}_{\mathrm{MOM}}$
depends on the direction of the 
momentum $p$ and on the coefficients $c_{\mu_1, \ldots ,\mu_n}$. 

For $\cO^S$ and $\cO^P$ we obtain 
\begin{equation}
Z^{\overline{\mathrm{MS}}}_{\mathrm{MOM}} =
    1 + \frac{g^2}{16 \pi^2} (4 + \xi) C_F + O(g^4) \,.
\end{equation}
In this special case we can go one step further and use the 
two-loop result for $Z^{\overline{\mathrm{MS}}}_{\mathrm{MOM}}$
\cite{fralu}.
For three colours and $n_f$ flavours one has in Landau gauge:
\begin{equation} \label{c2loop}
Z^{\overline{\mathrm{MS}}}_{\mathrm{MOM}} =
    1 + \frac{16}{3} \cdot \frac{g^2}{16 \pi^2} 
      + \left( 177.48452 - \frac{83}{9} n_f \right)
        \left( \frac{g^2}{16 \pi^2} \right)^2 + O(g^6) \,.
\end{equation}

We want to compare our nonperturbative results with
the corresponding values obtained in (tadpole improved) perturbation
theory on the lattice. For the renormalisation factor which brings 
us from the bare lattice operator to the 
renormalised operator in the $\overline{\mathrm{MS}}$ scheme,
lattice perturbation theory yields results of the form
\begin{equation} \label{zpert}
Z^{\mathrm{pert}}_\cO
 = 1 - \frac{g^2}{16 \pi^2} ( \gamma_0 \ln (a \mu) + 
          C_F \Delta_\cO ) 
\end{equation}
where $\Delta_\cO$ is a finite constant and $\gamma_0$ is the one-loop
coefficient of the anomalous dimension. Working with an anticommuting 
$\gamma_5$ also in the continuum part of the calculation we arrive
at the values given in Table~\ref{tab.finco} (see  Ref.~\cite{pertz} for more
details).

\begin{table}
\caption{Finite contributions to the renormalisation factors in lattice
         perturbation theory. In addition, the one- and two-loop 
         coefficients of the anomalous dimension are given for 0 flavours.}
\label{tab.finco}
\vspace{1.0cm}
\begin{center}
\begin{tabular}{crcr}
\hline
$\cO$ & \multicolumn{1}{c}{$\Delta_\cO$} & $\gamma_0$ 
      & \multicolumn{1}{c}{$\gamma_1$} \\
\hline
$\cO_{v_{2,a}} $ &     1.2796   &  64/9   &      96.69 \\
$\cO_{v_{2,b}} $ &     2.5619   &  64/9   &      96.69 \\
$\cO_{v_3} $     &   $-12.1274$ &  100/9  &     141.78 \\
$\cO_{v_4} $     &   $-27.2296$ &  628/45 &     172.58 \\
$\cO_{a_2} $     &   $-12.1171$ &  100/9  &     141.78 \\
$\cO_{r_{2,a}} $ &     0.3451   & 64/9    &      96.69 \\
$\cO_{r_{2,b}} $ &     0.1674   & 64/9    &      96.69 \\
$\cO_{r_3} $     &   $-12.8589$ & 100/9   &     141.78 \\
$\cO^S $         &    12.9524   &   $-8$  &  $-134.67$ \\
$\cO^P $         &    22.5954   &   $-8$  &  $-134.67$ \\
$\cO^V_\mu $     &    20.6178   &     0   &       0.00 \\
$\cO^A_\mu $     &    15.7963   &     0   &       0.00 \\
\hline
\end{tabular}
\end{center}
\vspace{1.0cm}
\end{table} 

In order to obtain the corresponding results in tadpole improved
perturbation theory \cite{lepmac} we write (with $\mu = 1/a$)
for an operator with $n_D$ covariant derivatives
\begin{eqnarray}
  1 - \frac{g^2}{16 \pi^2} C_F \Delta_\cO & = &
  \frac{u_0}{u_0^{n_D}} u_0^{n_D-1}  
           \left( 1 - \frac{g^2}{16 \pi^2} C_F \Delta_\cO \right) 
\nonumber \\
{} & = & \frac{u_0}{u_0^{n_D}} 
     \left( 1 - \frac{g^{*2}}{16 \pi^2} C_F \overline{\Delta}_\cO \right) 
       + O(g^{*4}) \,,
\end{eqnarray}
where
\begin{equation}
  u_0 = \langle \third \mathrm{tr} U_{\Box} \rangle ^{\frac{1}{4}} =
     1 - \frac{g^2}{16 \pi^2} C_F \pi^2 + O(g^4) 
\end{equation}
and
\begin{equation} 
  \overline{\Delta}_\cO = \Delta_\cO + (n_D - 1) \pi^2  \,.
\end{equation}
This reflects the fact that one has $n_D$ operator tadpole diagrams
and one leg tadpole diagram, which are of the same magnitude but
contribute with opposite sign. 
Furthermore, one chooses as the
expansion parameter $g^*$ the coupling constant renormalised at some
physical scale. We have taken $g^*$ from the values given for
$\alpha_{\overline{\mathrm{MS}}} (1/a)$ in Table~I of Ref.~\cite{lepmac}.

At this point we have two options. Either we stay with the expression
(\ref{zpert}) and its tadpole improved analogue  
\begin{equation} \label{zpertti}
  Z^{\mathrm{ti}}_\cO = u_0^{1-n_D} \left[
   1 - \frac{g^{*2}}{16 \pi^2} ( \gamma_0 \ln (a \mu) + 
          C_F \overline{\Delta}_\cO ) \right] 
\end{equation}
or we apply these formulae only at a fixed scale $\mu = \mu_0$ 
(e.g.\ $\mu_0 = 1/a$) using the
renormalisation group to change $\mu$. 

For this scale dependence, the renormalisation group predicts
(at fixed bare parameters)
\begin{equation} \label{rengroup}
 R_\cO (\mu,\mu_0) := \frac{Z_\cO (\mu)}{Z_\cO (\mu_0)} = 
  \exp \left \{ - \int_{\bar{g}(\mu_0^2)}^{\bar{g}(\mu^2)} 
    \mathrm{d}g \frac{\gamma(g)}{\beta(g)} \right \} \,.
\end{equation}
Here the $\beta$-function is given by
\begin{equation}
  \beta (g) = - \beta_0 \frac{g^3}{16 \pi^2} - 
             \beta_1 \frac{g^5}{(16 \pi^2)^2} + \cdots
\end{equation}
with
\begin{equation}
  \beta_0 = 11 - \frac{2}{3} n_f \,,\, \beta_1 = 102 - \frac{38}{3} n_f \,.
\end{equation}
In terms of the $\Lambda$ parameter the running coupling reads
\begin{equation} \label{runco}
  \frac{\bar{g}^2 (\mu^2)}{16 \pi^2} = 
       \frac{1}{\beta_0 \ln(\mu^2/\Lambda^2)} -
        \frac{\beta_1}{\beta_0^3} \frac{\ln \ln(\mu^2/\Lambda^2) }
                              {\ln^2(\mu^2/\Lambda^2)} + \cdots
\end{equation}
and the anomalous dimension is expanded as
\begin{equation}
  \gamma (g) =  \gamma_0 \frac{g^2}{16 \pi^2} + 
             \gamma_1 \left( \frac{g^2}{16 \pi^2} \right)^2 + \cdots
\end{equation}
For our operators the one- and two-loop coefficients $\gamma_0$ and 
$\gamma_1$ of the anomalous
dimension are known in the $\overline{\mathrm{MS}}$ scheme 
\cite{flora,nano}. The values actually used are listed in 
Table~\ref{tab.finco}. Within the two-loop approximation we obtain 
\begin{equation} \label{renfac}
 R_\cO (\mu,\mu_0) =
   \left( \frac{\bar{g}^2(\mu^2)}{\bar{g}^2(\mu_0^2)} \right)
                  ^{\frac{\gamma_0}{2 \beta_0}}
 \left( \frac{1 + \frac{\beta_1}{\beta_0} \frac{\bar{g}^2(\mu^2)}{16 \pi^2}}
        {1 + \frac{\beta_1}{\beta_0} \frac{\bar{g}^2(\mu_0^2)}{16 \pi^2}}
    \right)^{\half \left( (\gamma_1 / \beta_1) - 
              (\gamma_0 / \beta_0) \right)} \,.  
\end{equation}

In the case of the operators $\cO^S$ and $\cO^P$ we can do better and
use the three-loop expressions for the $\beta$ function and the anomalous
dimension. In the $\overline{\mathrm{MS}}$ scheme one finds
\cite{ritbergen,vermaseren,chetyrkin}
\begin{eqnarray}
 \beta_2 & = & \frac{2857}{2} - \frac{5033}{18} n_f 
           + \frac{325}{54} n_f^2 \,,  \nonumber \\ 
 \gamma_1 & = & - \frac{404}{3} + \frac{40}{9} n_f \,, \\
 \gamma_2 & = & - 2498 + 
            \left( \frac{4432}{27} + \frac{320}{3} \zeta_3 \right) n_f
            + \frac{280}{81} n_f^2 \,, \nonumber
\end{eqnarray}
where $\zeta_3 \approx 1.2020569032$. For the running coupling we have now
instead of (\ref{runco})
\begin{eqnarray} 
  \frac{\bar{g}^2 (\mu^2)}{16 \pi^2} & = & 
       \frac{1}{\beta_0 \ln(\mu^2/\Lambda^2)} -
        \frac{\beta_1}{\beta_0^3} \frac{\ln \ln(\mu^2/\Lambda^2) }
                              {\ln^2(\mu^2/\Lambda^2)}  
    \nonumber \\ 
 {} & {} & {} + \frac{1}{\beta_0^5 \ln^3(\mu^2/\Lambda^2)}
    \left( \beta_1^2 \ln^2 \ln(\mu^2/\Lambda^2)
           - \beta_1^2 \ln \ln(\mu^2/\Lambda^2) \right. 
    \nonumber \\   
{} & {} & \hspace{5.0cm} {} + \left. \beta_2 \beta_0 - \beta_1^2 \right) 
                            + \cdots
\end{eqnarray}
and $R_\cO$ takes the form 
\begin{equation} 
 R_\cO (\mu,\mu_0) = 
  \frac{\exp F \left(\frac{\bar{g}^2(\mu^2)}{16 \pi^2}\right)}
       {\exp F \left(\frac{\bar{g}^2(\mu_0^2)}{16 \pi^2}\right)}
\end{equation}
with
\begin{eqnarray} 
  F(x) & = & \frac{\gamma_0}{2 \beta_0} \ln x
     + \frac{\beta_0 \gamma_2 - \beta_2 \gamma_0}{4 \beta_0 \beta_2}
        \ln \left(\beta_0 + \beta_1 x + \beta_2 x^2 \right)
   \nonumber \\ 
  {} & {} & {}  + \frac{2 \beta_0 \beta_2 \gamma_1 
        - \beta_1 \beta_2 \gamma_0 - \beta_0 \beta_1 \gamma_2 }    
            {2 \beta_0 \beta_2 \sqrt{4 \beta_0 \beta_2 - \beta_1^2}}
       \arctan \left( \frac{\beta_1 + 2 \beta_2 x}
                        {\sqrt{4 \beta_0 \beta_2 - \beta_1^2}} \right) \,.
\end{eqnarray}

Having data at two different values of the bare coupling we can
also compare the dependence on the bare coupling with the behaviour
expected from the renormalisation group. Keeping the renormalisation
scale $\mu$ and the renormalised quantities fixed we find for
the ratio of the renormalisation factors $Z_\cO$ and $Z_\cO^\prime$ 
at the bare couplings $g$ and $g^\prime$, respectively,
\begin{equation}
 \frac{Z_\cO}{Z_\cO^\prime} = 
  \exp \left \{ \int_{g^\prime}^{g} 
    \mathrm{d}g_0 \frac{\hat{\gamma}(g_0)}{\hat{\beta}(g_0)} \right \} \,.
\end{equation}
Here $\hat{\gamma}(g)$ and $\hat{\beta}(g)$ denote the anomalous
dimension and the $\beta$-function obtained by differentiation with
respect to the cut-off at fixed renormalised quantities. They are
considered as functions of the bare coupling.
Within the two-loop approximation we get
\begin{equation} \label{renfacb}
 \frac{Z_\cO}{Z_\cO^\prime} = 
   \left( \frac{g^2}{g^{\prime 2}} \right)
                  ^{-\frac{\gamma_0}{2 \beta_0}}
 \left( \frac{1 + \frac{\beta_1}{\beta_0} \frac{g^2}{16 \pi^2}}
        {1 + \frac{\beta_1}{\beta_0} \frac{g^{\prime 2}}{16 \pi^2}}
    \right)^{-\half \left( (\hat{\gamma}_1 / \beta_1) - 
              (\gamma_0 / \beta_0) \right)} \,,
\end{equation}
where we have used the fact that $\beta_0$, $\beta_1$, and $\gamma_0$
are universal so that $\hat{\beta}_0 = \beta_0$, $\hat{\beta}_1 = \beta_1$,
$\hat{\gamma}_0 = \gamma_0$. The coefficient $\hat{\gamma}_1$ reads
(cf.\ also \cite{alles})
\begin{equation}
 \hat{\gamma}_1 = \gamma_1 + 2 \beta_0 C_F \Delta_\cO
    - 32 \pi^2 \gamma_0 \left( - 0.234101
      + n_f \cdot 0.0034435 \right)  
\end{equation}
for the gauge group SU(3). The quantities $\Delta_\cO$ are
given in Table~\ref{tab.finco}, and the coefficient of $\gamma_0$ stems
from the ratio of the $\Lambda$-parameters on the lattice and in the 
$\overline{\mathrm{MS}}$ scheme (see, e.g., \cite{momue}). 

\section{Numerical implementation} \label{sec.numerics}

Let us sketch the main ingredients of our calculational procedure.
To simplify the notation we set the lattice spacing $a=1$ in this section.
In a first step gauge field configurations are generated and numerically
fixed to some convenient gauge (the Landau gauge in our case). 
The non-amputated Green function (\ref{Gdef}) is calculated as the 
gauge field average of $\hat{G} (p)$, which is constructed from the quark 
propagator $\hat{S} (p)$ on the same gauge field configuration 
according to
\begin{equation} \label{greenf}
 \hat{G}(p) = \frac{1}{V}
   \sum_{x,y,z,z^\prime} {\mathrm e}^{- {\mathrm i} p \cdot (x-y) } 
    \hat{S}(x,z) J(z,z^\prime) \hat{S}(z^\prime,y) \,.
\end{equation}
We omit the quark-line disconnected contribution (which would 
contribute only to flavour-singlet operators) and suppress
Dirac as well as colour indices. The operator under study is
represented by $J(z,z^\prime)$:
\begin{equation}
\sum_z \cO (z) = \sum_{z,z^\prime} \bar{q}(z) J(z,z^\prime) q(z^\prime) \,.
\end{equation}
Using the relation
\begin{equation}
 \hat{S}(x,y) = \gamma_5 \hat{S}(y,x)^+ \gamma_5
\end{equation}
we rewrite $\hat{G}(p)$ as  
\begin{equation}
 \hat{G}(p) = \sum_{z,z^\prime} \gamma_5 
  \left( \sum_x \hat{S}(z,x) {\mathrm e}^{{\mathrm i} p \cdot x} \right)^+ 
  \gamma_5 J(z,z^\prime)
  \left( \sum_y \hat{S}(z^\prime,y) {\mathrm e}^{{\mathrm i} p \cdot y} \right)
\end{equation}
in terms of the quantities
\begin{equation}
   \sum_x \hat{S}(z,x) {\mathrm e}^{{\mathrm i} p \cdot x} \,.
\end{equation}
These can be calculated by solving the lattice Dirac equation 
with a momentum source:
\begin{equation}
 \sum_z M(y,z)   
   \left( \sum_x \hat{S}(z,x) {\mathrm e}^{{\mathrm i} p \cdot x} \right) 
   =  {\mathrm e}^{{\mathrm i} p \cdot y} \,.
\end{equation}
Here $M(x,y)$ represents the fermion matrix. So the number of matrix
inversions to be performed is proportional to the number of momenta
considered. But the quark propagators, which we need for the amputation
and the computation of the quark wave function renormalisation, are 
immediately obtained from the quantities already calculated.

Another computational strategy would be to choose a particular 
location for the operator, i.e.\ instead of summing over $z^\prime$
(and $z$) in (\ref{greenf}) one could set $z^\prime = 0$. 
Translational invariance
tells us that this will give the same expectation value after
averaging over all gauge field configurations. For this method
we need to solve the Dirac equation with a point source at the
location of the operator and (in the case of extended operators)
for a small number of point sources in the immediate neighbourhood.
For operators with a small number of derivatives the point source
method would require fewer inversions, but it turns out that
relying on translational invariance leads to much larger statistical 
errors. This is clearly seen in the results shown in Ref.~\cite{melbourne},
which were obtained with a point source instead of a momentum source
(cf.\ also Ref.~\cite{harald}). 

We work with standard Wilson fermions ($r=1$) in the quenched 
approximation. At $\beta = 6.0$ we have analysed 20 configurations
on a $16^3 \times 32$ lattice for three values of the hopping parameter,
$\kappa = 0.155$, 0.153, and 0.1515. 
Four further configurations at
$\beta = 6.2$ on a $24^3 \times 48$ lattice were studied with
$\kappa = 0.152$, 0.1507, and 0.1489.
For the critical $\kappa$ we take the values $\kappa_c = 0.157211$
($\beta = 6.0$) and $\kappa_c = 0.153374$ ($\beta = 6.2$), respectively
(see Ref.~\cite{scaling}).

Before we calculate quark correlation functions on our configurations
we fix the gauge to the Landau gauge \cite{mandula}.
The gauge fixing necessarily raises the question of the influence of
Gribov copies. Fortunately, an investigation of this problem in a
similar setting indicates that the fluctuations induced by the Gribov
copies are not overwhelmingly large and may be less important than the
ordinary statistical fluctuations \cite{gribcop}. Therefore we decided
to neglect the Gribov problem for the time being, but a careful study
in the case at hand is certainly desirable. 

Coming finally to the choice of the momenta, we tried to avoid
momenta along the coordinate axes in order to minimise cut-off effects.
Altogether we used 43 momenta at $\beta = 6.0$ and 41 momenta
at $\beta = 6.2$.

\section{Results} \label{sec.results}

For the presentation of our results we convert lattice units 
to physical units using $a^{-2} = 3.8$ GeV$^2$ ($\beta = 6.0$) 
and $a^{-2} = 7.0$ GeV$^2$ ($\beta = 6.2$)
as determined from the string tension \cite{scaling}. This leads to
values for the spatial extent of our lattices of 1.6~fm 
($\beta = 6.0$) and 1.8~fm ($\beta = 6.2$).
The $\Lambda$ parameter 
is taken to be $\Lambda_{\overline{\mathrm{MS}}} = 230$ MeV. 
This value follows from the result given in Ref.~\cite{alpha} by
using the string tension instead of the force parameter $r_0$ 
to set the scale. As we work in the quenched approximation we 
have to put $n_f = 0$ in the perturbative formulae.
The plotted errors are purely statistical and have been calculated
by a jackknife procedure.

\begin{figure}
\vspace*{-2.0cm}
\epsfig{file=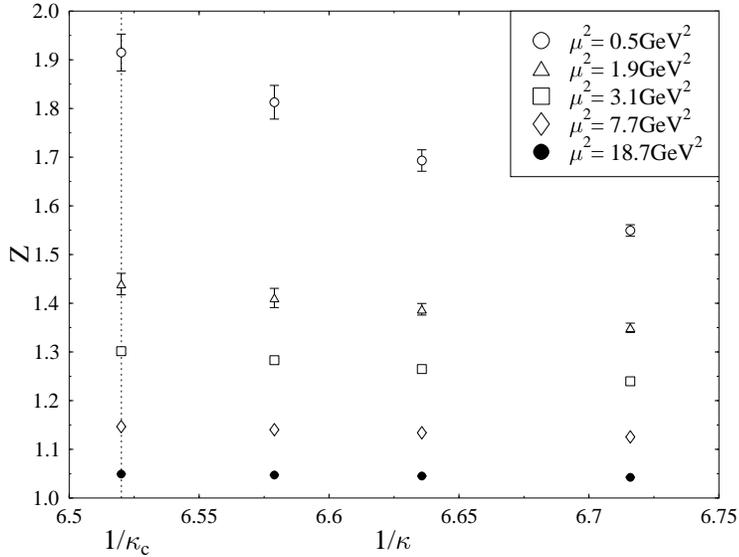,width=13cm} 
\vspace*{-1.0cm}
\caption{Chiral extrapolation of $Z$ for $\cO_{v_{2,a}}$ at $\beta = 6.2$.
         Some representative values of $\mu^2$ have been selected.}
\label{fig.chiex}
\end{figure}

The quark masses used in our simulations are rather large. 
On the other hand, the perturbative calculations, which we use
for comparison and for converting to the 
$\overline{\mathrm{MS}}$ scheme, are performed at
vanishing quark mass. Hence we first extrapolate our results to
the chiral limit. This is done linearly in $1/\kappa$. In most cases
the mass dependence is quite weak so that the extrapolation looks 
reliable, see Fig.~\ref{fig.chiex} for an example (operator 
$\cO_{v_{2,a}}$ at $\beta = 6.2$). 
This is not too surprising
since the mass scale in the calculations is not set by the quark 
mass, but by the (off-shell) momentum, which is typically
much larger. The only exception is the pseudoscalar density where
a rather strong mass dependence is found. This observation already 
indicates that the pseudoscalar density is something special, and
we will discuss this operator in more detail in 
Subsection~\ref{subsec.density}.

\begin{figure}
\vspace*{-2.0cm}
\epsfig{file=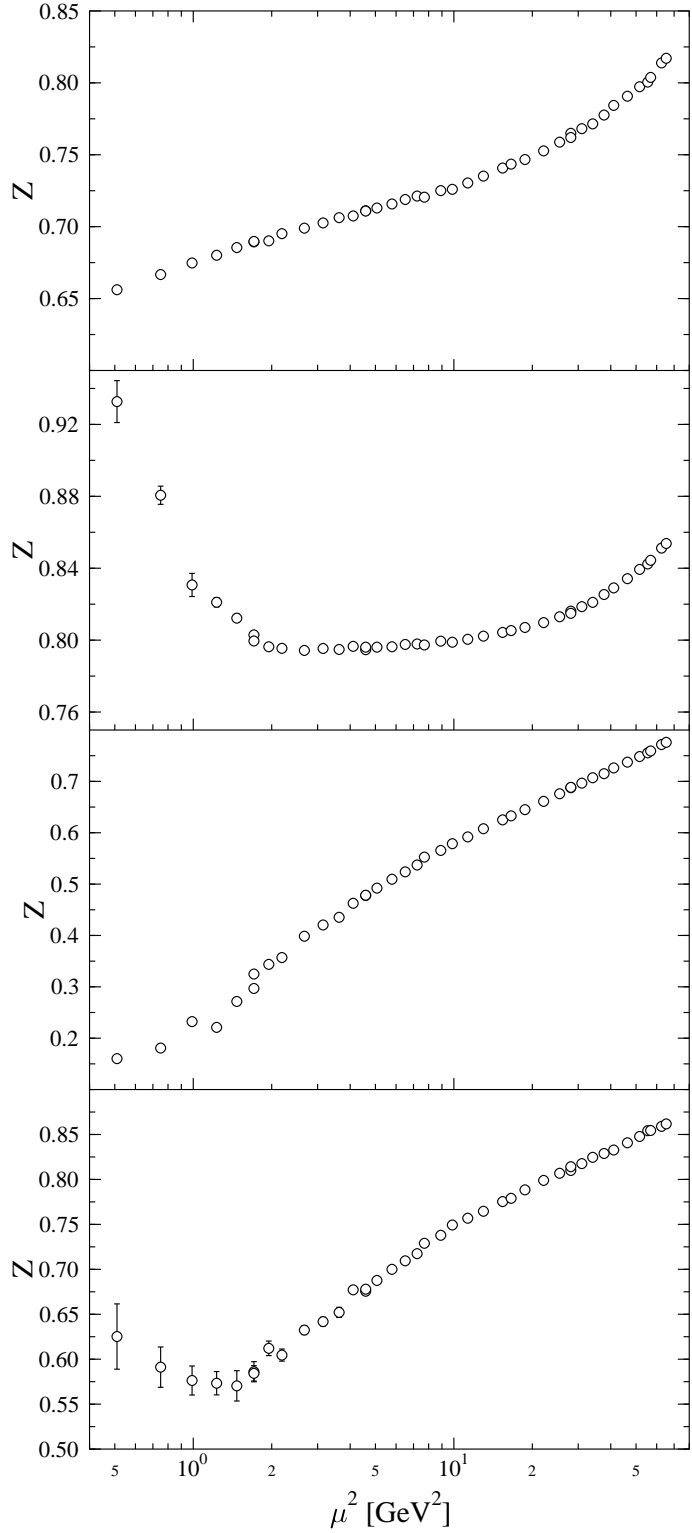,width=16cm} 
\caption{$Z$ in the MOM scheme for $\cO^V_\mu$, $\cO^A_\mu$, $\cO^P$,
         and $\cO^S$ (from top to bottom) at $\beta = 6.2$. 
         For the local vector current
         and the axial vector current, $Z$ has been determined from
         the transverse components (cf.\ Section~\ref{sec.vector}).}
\label{fig.zmom.I}
\end{figure}
\begin{figure}
\vspace*{-2.0cm}
\epsfig{file=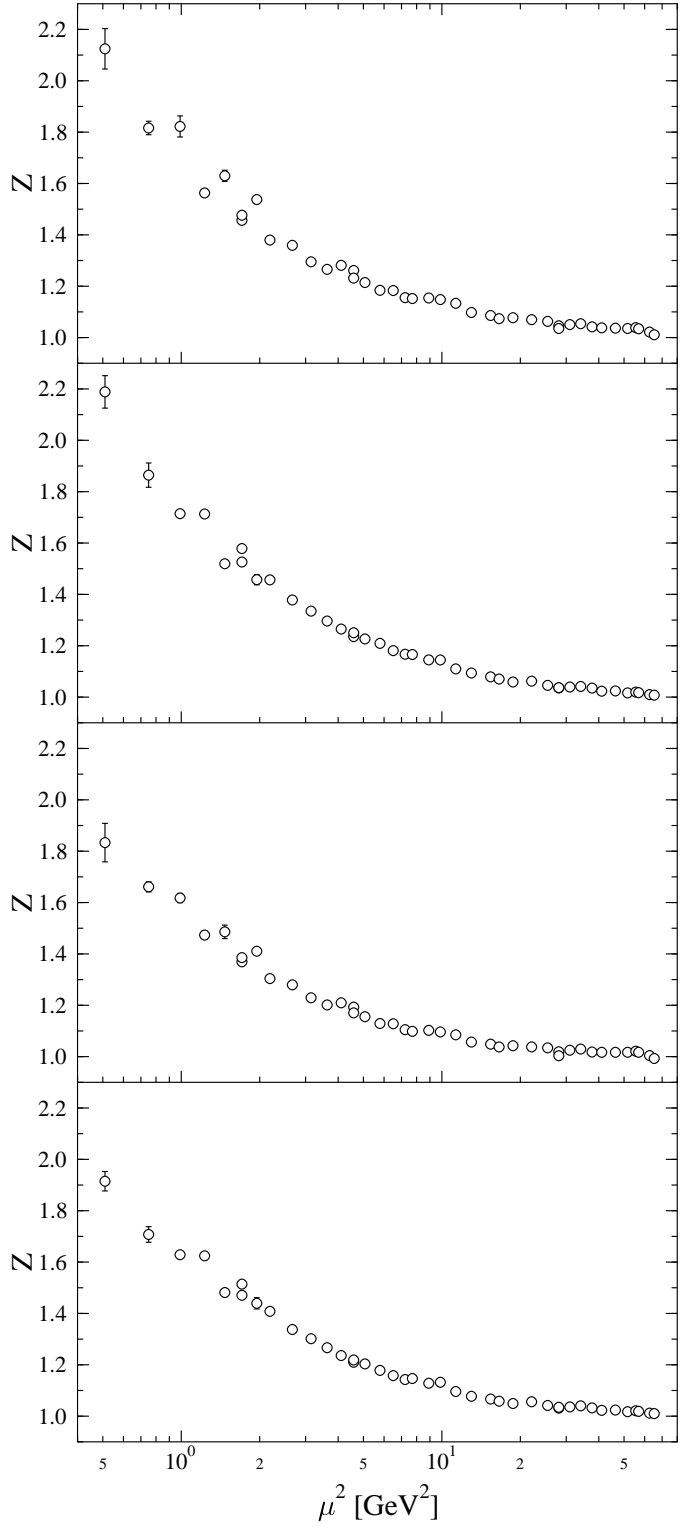,width=16cm} 
\caption{$Z$ in the MOM scheme for $\cO_{r_{2,b}}$, $\cO_{r_{2,a}}$, 
         $\cO_{v_{2,b}}$, and $\cO_{v_{2,a}}$ (from top to bottom)
         at $\beta = 6.2$.}
\label{fig.zmom.v2a}
\end{figure}
\begin{figure}
\vspace*{-2.0cm}
\epsfig{file=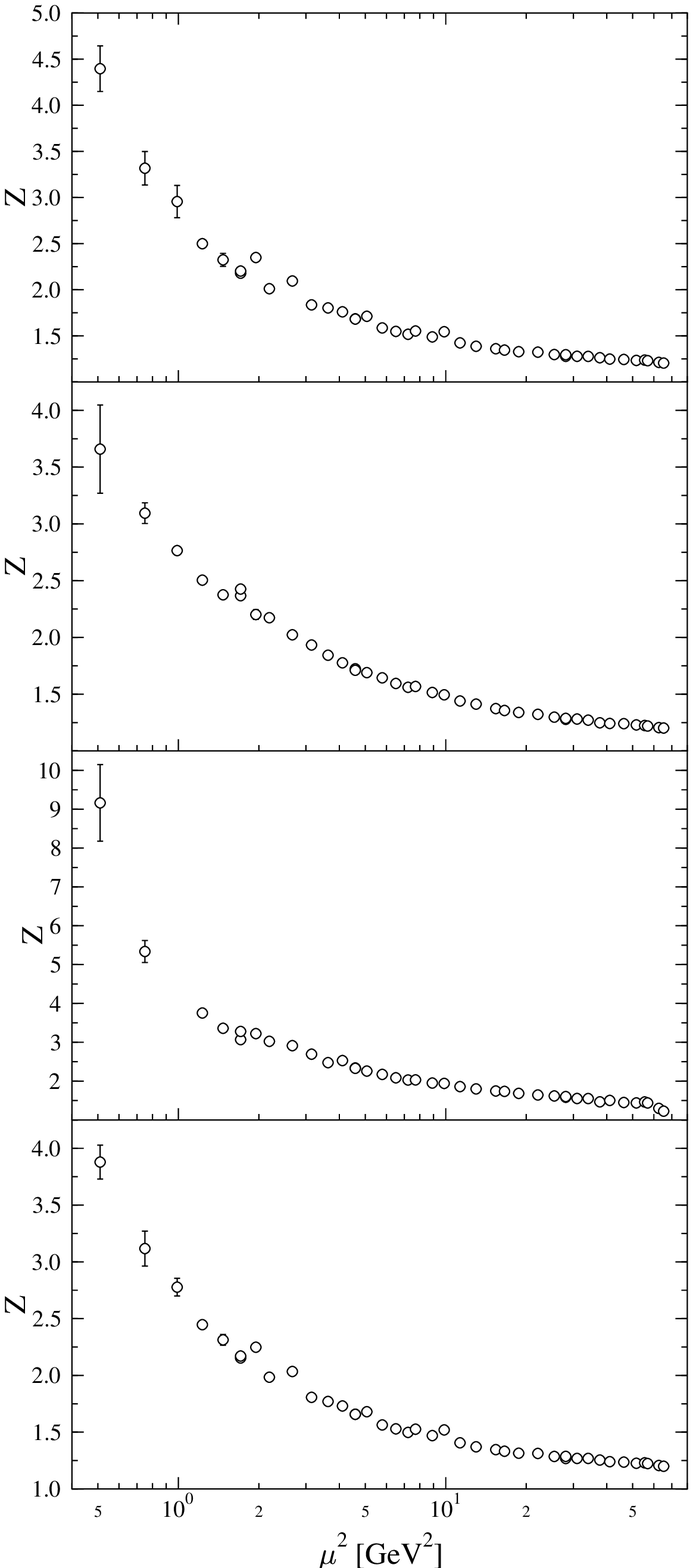,width=16cm} 
\caption{$Z$ in the MOM scheme for $\cO_{r_3}$, $\cO_{a_2}$, 
         $\cO_{v_4}$, and $\cO_{v_3}$ (from top to bottom)
         at $\beta = 6.2$.}
\label{fig.zmom.v3}
\end{figure}

\begin{figure}
\vspace*{-2.0cm}
\epsfig{file=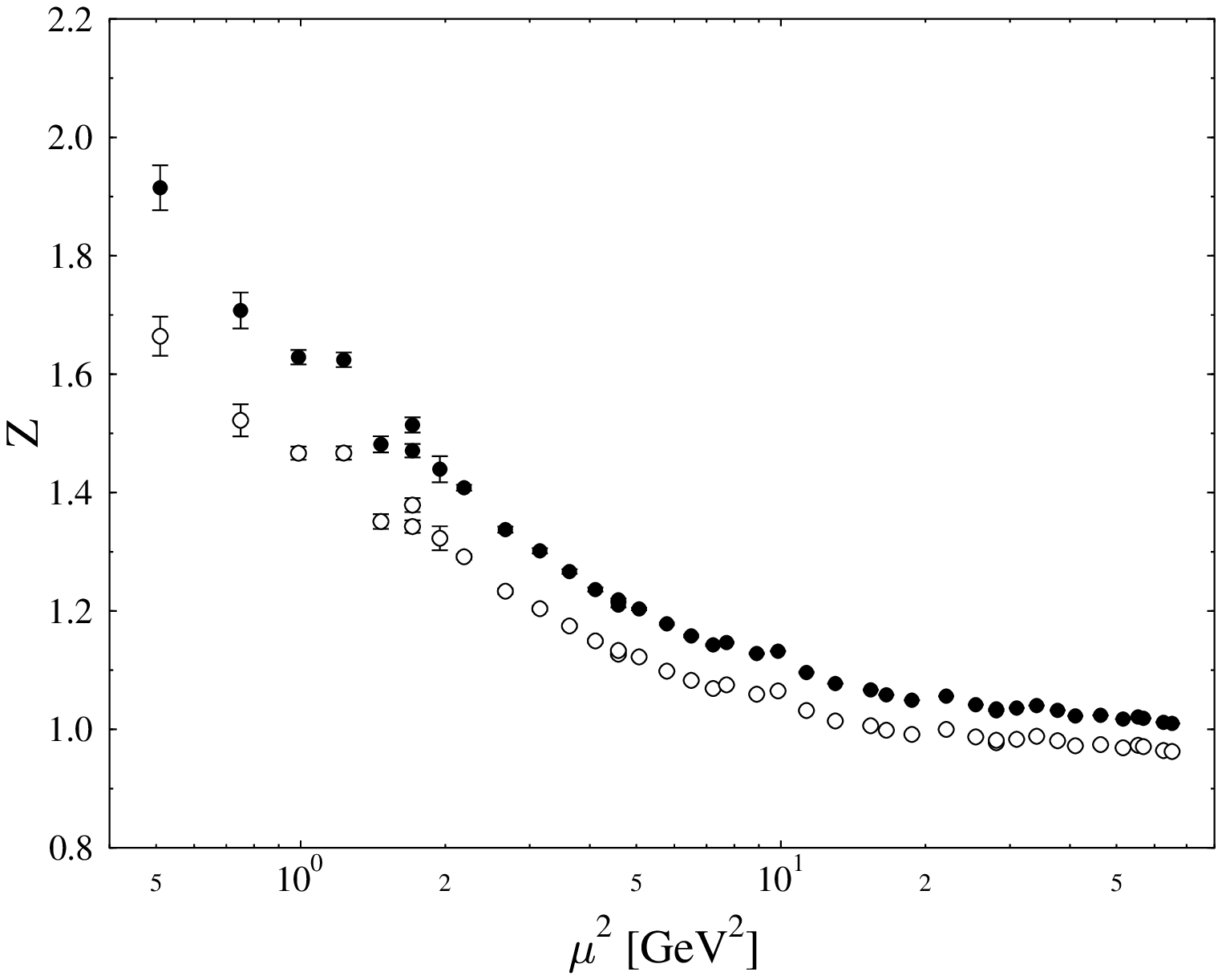,width=13cm} 
\vspace*{-1.0cm}
\caption{$Z$ for the operator $\cO_{v_{2,a}}$ at $\beta = 6.2$. 
         The filled circles represent the data corresponding to the 
         MOM renormalisation condition (\ref{defz}), the open circles
         represent the values in the  $\overline{\mathrm{MS}}$ scheme.}
\label{fig.v2a.mom-ms}
\end{figure}

In Figs.~\ref{fig.zmom.I} - \ref{fig.zmom.v3} we display our 
``raw'' results, i.e.\ the nonperturbative
$Z$'s in our MOM scheme, for $\beta = 6.2$. 
Then we multiply by $Z^{\overline{\mathrm{MS}}}_{\mathrm{MOM}}$
in order to obtain the $Z$ factors in the $\overline{\mathrm{MS}}$ scheme.
When evaluating the perturbative expression for 
$Z^{\overline{\mathrm{MS}}}_{\mathrm{MOM}}$
we insert for the coupling constant the running coupling in the 
$\overline{\mathrm{MS}}$ scheme taken at the renormalisation scale $\mu^2$
(cf.\ Eq.~(\ref{runco})). 
Fig.~\ref{fig.v2a.mom-ms} illustrates the effect of the conversion to the
$\overline{\mathrm{MS}}$ scheme for the operator
$\cO_{v_{2,a}}$ at $\beta = 6.2$.

Ultimately we are interested in physical observables, e.g.\ in 
the moments of the structure functions. They are obtained
from the bare matrix elements calculated on the lattice after
multiplication by the appropriate $Z$ factor and the Wilson
coefficient (both computed in the $\overline{\mathrm{MS}}$ scheme).
So the bridge from the lattice to continuum physics
consists of the product of these two quantities. Having calculated $Z$
nonperturbatively we can multiply with the corresponding Wilson
coefficient in order to see if the expected cancellation of the $\mu$
dependence really occurs. If it does, the $\mu$ independent value of
the product is the factor we need in order to calculate the moments of the
structure functions from the lattice results. Since this cancellation
means that the scale dependence of our $Z$'s should be described by the
renormalisation group factor $R$ (given by (\ref{renfac}) 
in two-loop approximation)
we shall divide our numerical results by this expression
and define $Z_{\mathrm{RGI}} = Z/R$. For $Z_{\mathrm{RGI}}$ 
we hope to obtain a $\mu$ independent answer, at least in a reasonable
window of $\mu$ values. We shall choose $\mu_0^2 = 4$ GeV$^2$
in (\ref{renfac}).

\subsection{Vector and axial vector currents} \label{subsec.veccur}

\begin{figure}
\vspace*{-1.0cm}
\epsfig{file=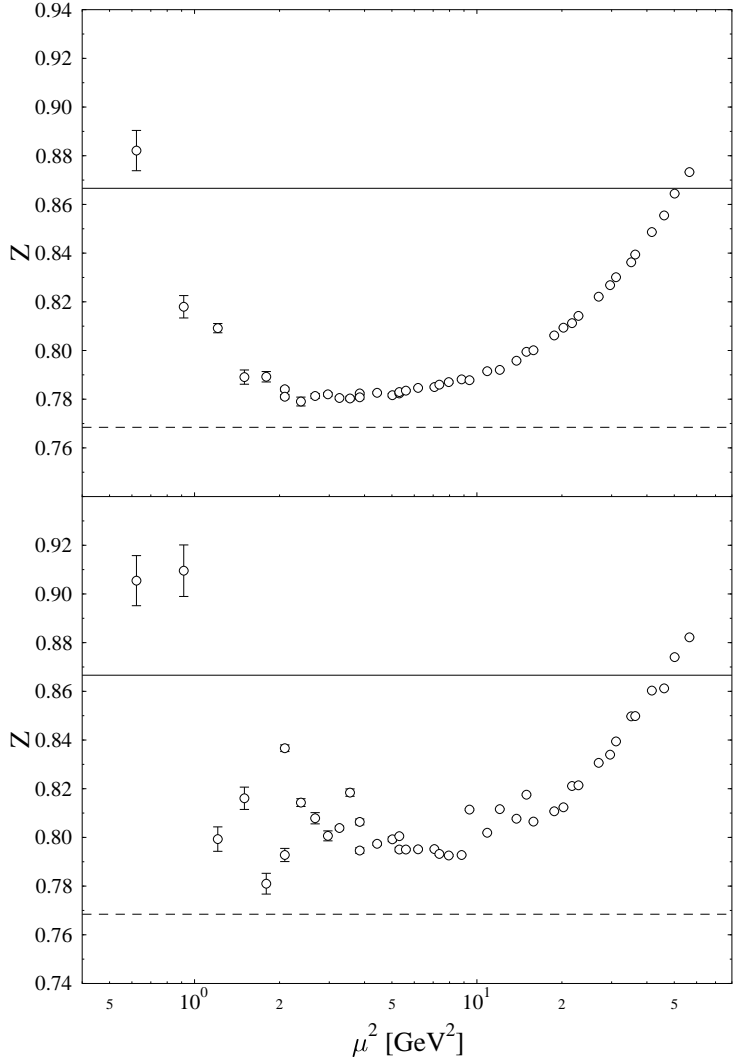,width=13cm} 
\caption{$Z$ for the axial vector current (transverse component)
         at $\beta = 6.0$ determined by the method of 
         Section~\ref{sec.vector} (upper plot).
         The lower plot shows $Z$ for the operator 
         $\bar{q} \gamma_4 \gamma_5 q$ at the same $\beta$ value. 
         The dashed (solid) line represents the prediction of
         one-loop lattice perturbation theory with (without)
         tadpole improvement.}
\label{fig.va0-vah}
\end{figure}

Let us begin with the vector and axial vector currents. Since in
these cases the $\overline{\mathrm{MS}}$ anomalous dimensions vanish, 
the renormalisation group factor (\ref{renfac}) equals 1 and 
we expect to find immediately scale independent results in a 
suitable window. As already mentioned, the 
standard procedure from Section~\ref{sec.method} suffers 
more strongly from lattice artifacts than the alternative 
method described in Section~\ref{sec.vector}.
This is exemplified in Fig.~\ref{fig.va0-vah}.
The results obtained by applying the standard procedure to
$\bar{q} \gamma_4 \gamma_5 q$ scatter more strongly than the $Z$'s
from the transverse component of the axial vector current.
In the remainder of this paper we shall only use 
the values delivered by the latter procedure.

Fig.~\ref{fig.va0-vah} nicely fulfills our expectations. We see a 
flat region around $\mu^2 \approx 4$ GeV$^2$ where $Z$ becomes
scale independent as it should.
Moreover, tadpole improvement has really improved 
the agreement between our nonperturbative result and one-loop 
lattice perturbation theory. 

\begin{figure}
\vspace*{-2.0cm}
\epsfig{file=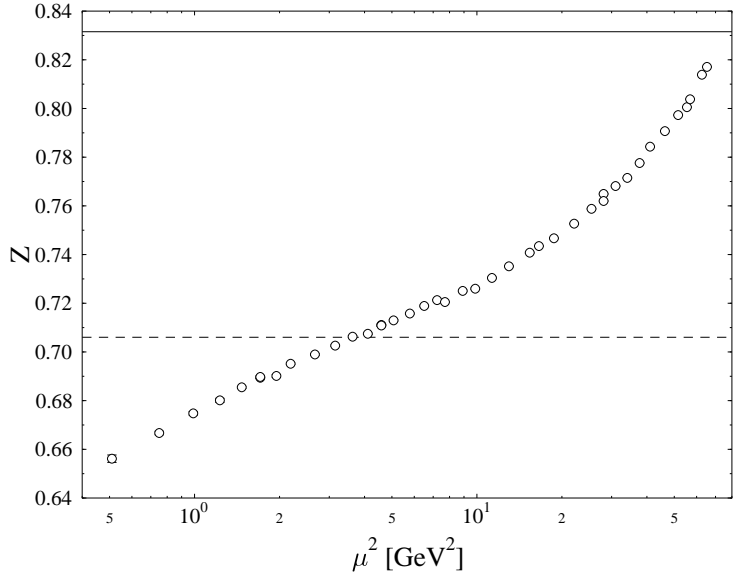,width=13cm} 
\vspace*{-1.0cm}
\caption{$Z$ for the local vector current (transverse component)
         at $\beta = 6.2$ determined by the method of 
         Section~\ref{sec.vector}. 
         The dashed (solid) line represents the prediction of
         one-loop lattice perturbation theory with (without)
         tadpole improvement.}
\label{fig.vlh}
\end{figure}

The analogous plot for the transverse component of the local 
vector current is displayed in
Fig.~\ref{fig.vlh}. No indication of a scaling window is seen.
How can this different behaviour of local vector and axial vector
current be explained? A hint may be obtained from lattice perturbation
theory. If one keeps all terms of order $am$ one gets an
additional contribution of the qualitatively correct form in the 
case of the local vector current, whereas the analogous contribution
to the renormalisation of the axial vector current is considerably
smaller \cite{imppert}. Due to the spontaneous breakdown of
chiral symmetry one should expect that some mass-like effects
survive even in the chiral limit and may lead to the observed
behaviour of the local vector current through such $am$ terms.

\begin{figure}
\vspace*{-1.0cm}
\epsfig{file=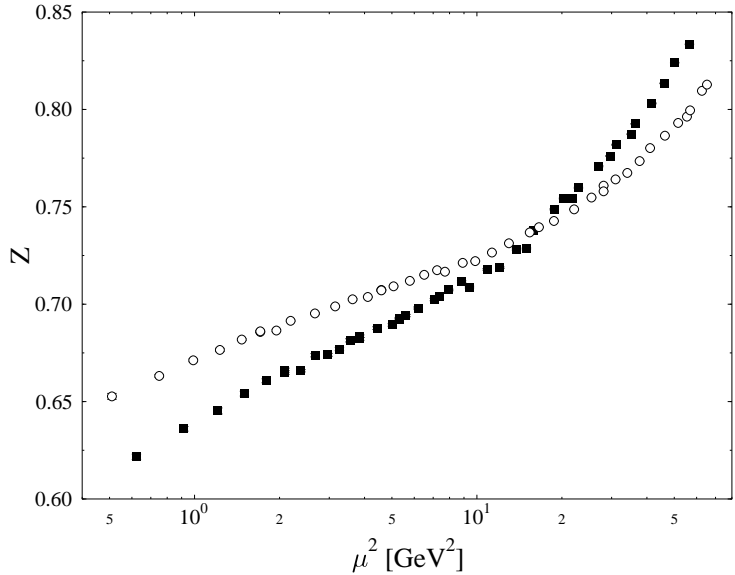,width=13cm} 
\vspace*{-1.0cm}
\caption{$Z$ for the local vector current (transverse component).
         The open circles (filled squares) 
         represent the data for $\beta = 6.2$ ($\beta = 6.0$).
         The $\beta = 6.2$ data have been rescaled perturbatively
         to $\beta = 6.0$ by multiplication with (\ref{renfacb}).} 
\label{fig.z2beta.vlh}  
\end{figure}

\begin{figure}
\vspace*{-2.0cm}
\epsfig{file=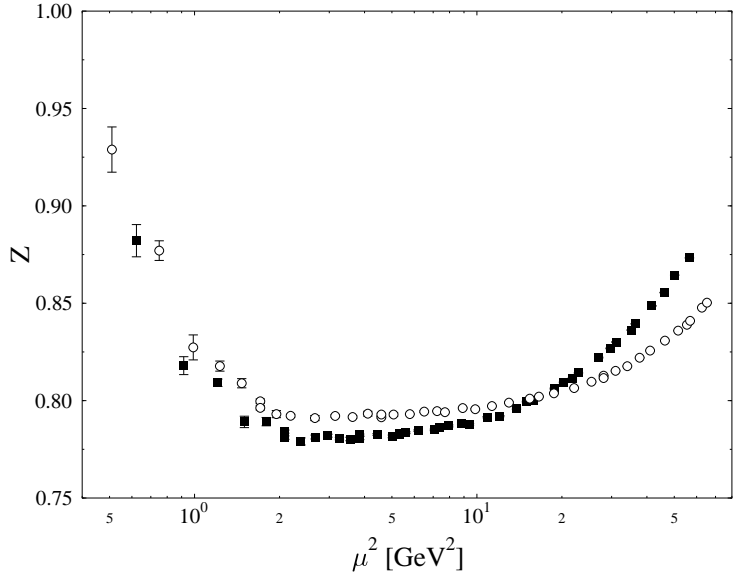,width=13cm} 
\vspace*{-1.0cm}
\caption{$Z$ for the axial vector current (transverse component).
         The open circles (filled squares) 
         represent the data for $\beta = 6.2$ ($\beta = 6.0$).
         The $\beta = 6.2$ data have been rescaled perturbatively
         to $\beta = 6.0$ by multiplication with (\ref{renfacb}).} 
\label{fig.z2beta.vah}  
\end{figure}

This interpretation is supported by a comparison of the data 
at $\beta = 6.0$ and $\beta = 6.2$.
The perturbative renormalisation group tells us that
the $\beta$ dependence
of the $Z$'s is given by the $\mu^2$ independent factor 
(\ref{renfacb}) in terms of the bare coupling constants
$g$ and $g^\prime$. Therefore we multiply the $\beta = 6.2$ data
with the factor (\ref{renfacb}) and plot them together with
the $Z$'s at $\beta = 6.0$ versus the renormalisation scale $\mu^2$.
The results are shown in Fig.~\ref{fig.z2beta.vlh} for the local 
vector current and in Fig.~\ref{fig.z2beta.vah} for the 
axial vector current. In both cases, the 
$\beta = 6.2$ data are closer to our expectations, but the local
vector current seems to be more sensitive to the variation of $a$
than the axial current, in accord with the above mentioned 
stronger influence of $am$ corrections in lattice perturbation theory.
Indeed, comparing the slopes of the data for the local vector
current at the two $\beta$ values in the region below 
$\mu^2 \approx 15$ GeV$^2$ one can estimate a ratio of about 1.45,
close to the ratio $a(\beta = 6.0)/a(\beta = 6.2) \approx 1.36$.

\subsection{Scalar and pseudoscalar density} \label{subsec.density}

\begin{figure}
\epsfig{file=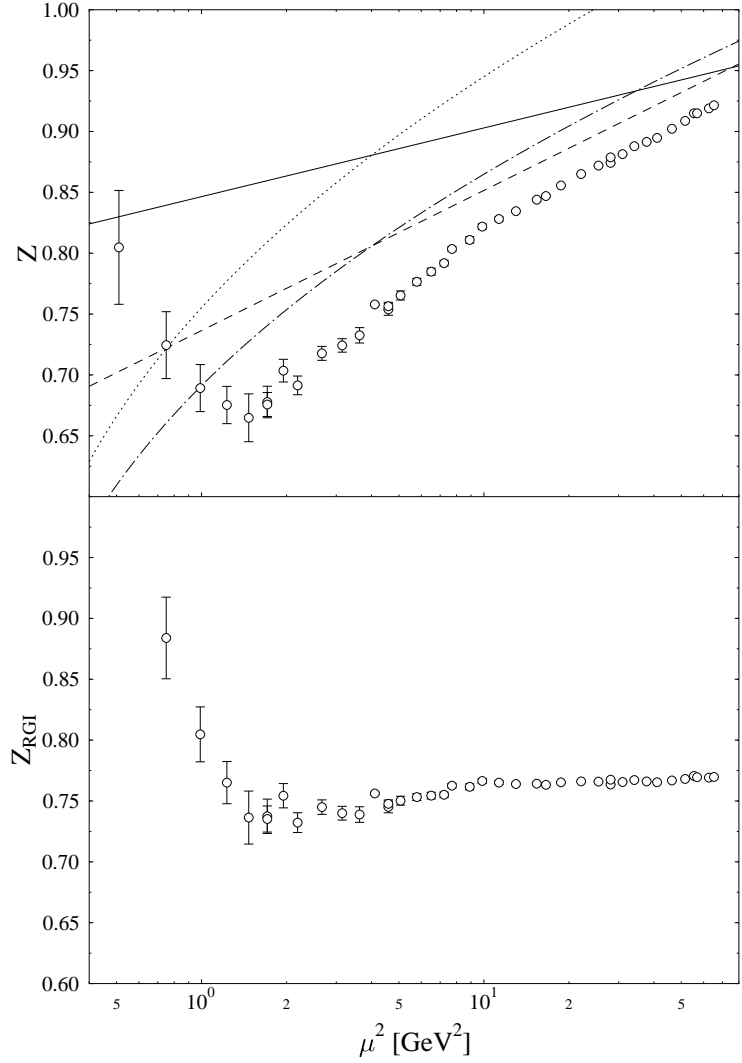,width=13cm} 
\caption{$Z$ and $Z_{\mathrm{RGI}}$ for the scalar density 
         at $\beta = 6.2$. 
         The dashed (solid) straight line represents the prediction of
         one-loop lattice perturbation theory with (without)
         tadpole improvement 
         (Eqs.\ (\ref{zpertti}), (\ref{zpert}), respectively). 
         The dotted curve results from improving the perturbative
         prediction with the renormalisation group (cf.\ Eq.\ 
         (\ref{rengroup})) and matching
         at $\mu_0^2 = 4$ GeV$^2$. The dash-dotted curve represents
         the same modification for tadpole improved
         perturbation theory.}
\label{fig.I.2pl}
\end{figure}

Fig.~\ref{fig.I.2pl} shows our results for the $Z$ factor of the 
scalar density.
The upper part of the figure compares the nonperturbative numbers
transformed to the $\overline{\mathrm{MS}}$ scheme with the predictions
from lattice perturbation theory with and without tadpole improvement.
We also display the curves resulting from renormalisation group
improvement applied to the perturbative results at $\mu_0^2 = 4$ GeV$^2$.
Tadpole improvement works in the right direction and the data seem to
follow the trend required by the renormalisation group 
if $\mu^2$ is not too small. This is more
clearly exhibited in the lower part of the figure where we have
divided by the renormalisation group factor (\ref{renfac}).
As expected, we find a nice scaling window in $Z_{\mathrm{RGI}}$ 
although some lattice artifacts are clearly visible. 

\begin{figure}
\vspace*{-2.0cm}
\epsfig{file=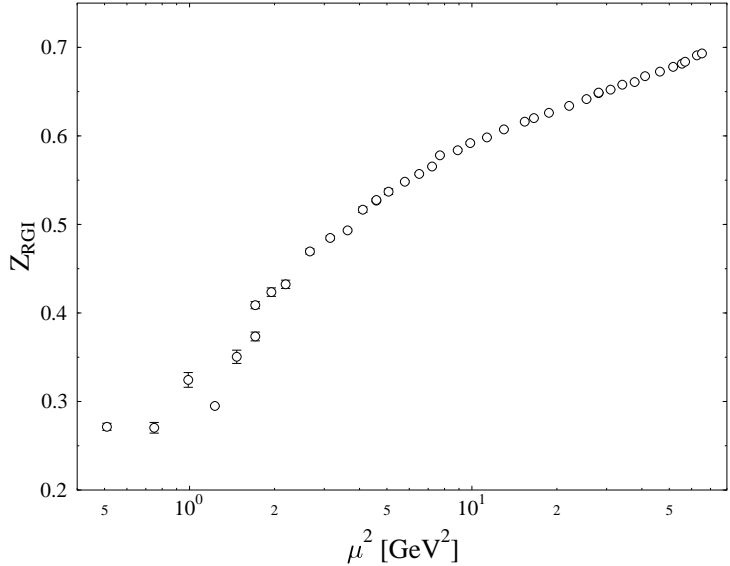,width=13cm} 
\vspace*{-1.0cm}
\caption{$Z_{\mathrm{RGI}}$ for the pseudoscalar density at $\beta = 6.2$.} 
\label{fig.r.g5}
\end{figure}

The results for the pseudoscalar density do not agree with the naive
expectation that $Z_{\mathrm{RGI}}$ should be constant
(see Fig.~\ref{fig.r.g5}). This observation was already made
by Martinelli et al.\cite{marti}. Moreover the data show a strong 
mass dependence, especially for the lower values of $\mu^2$. 
These features can be explained by the dynamics of chiral 
symmetry breaking \cite{marti}.
Indeed, in the continuum there is a Ward identity relating the
(bare) pseudoscalar vertex $\Gamma^P (p)$ at momentum transfer zero
to the quark propagator:
\begin{equation} \label{ward5}
 - 2 m_0 \Gamma^P (p) = S^{-1} (p) \gamma_5 + \gamma_5 S^{-1} (p) \,.
\end{equation}
Here $m_0$ is the bare quark mass, which is taken to be nonzero.
According to Pagels \cite{pagels}, spontaneous breakdown of chiral
symmetry produces a mass-like contribution to $S^{-1}$ which commutes 
(rather than anticommutes) with $\gamma_5$ and persists in the chiral limit.
Thus the r.h.s.\ of (\ref{ward5}) will not vanish for $m_0 \to 0$ 
and the pseudoscalar vertex must diverge. Consequently, the 
pseudoscalar density is ill defined if the chiral limit is performed
after the momentum transfer has been set to zero, which is the order
of limits in our procedure.

For our $Z$ factors the identity (\ref{ward5}) would lead to
\begin{equation} 
Z_{\cO^P} (p) = - \frac{12 m_0 Z_q (p)}{\mathrm{tr} S^{-1}(p)}
\end{equation}
and with $\mathrm{tr} S^{-1}(p)$ remaining nonzero even in the chiral
limit $Z_{\cO^P} (p)$ should vanish for $m_0 \to 0$. This tendency is
indeed shown by our data (at least for the lower values of $\mu^2$).

Nevertheless, for sufficiently large $\mu^2$, i.e.\ in the
perturbative regime, the ratio $Z_{\cO^P}/Z_{\cO^S}$ should
become constant, because $\cO^P$ and $\cO^S$ have equal
anomalous dimensions. Comparison of the lower part of 
Fig.~\ref{fig.I.2pl} with Fig.~\ref{fig.r.g5} shows that our data
do not display this behaviour. The ratio continues to rise
up to our largest values of $\mu^2$.

\subsection{Operators with derivatives}

In this subsection we want to discuss some selected operators
with derivatives, which determine moments of hadronic structure 
functions. Again, we would like to find a ``window'' 
at moderate values of $\mu^2$ where $Z_{\mathrm{RGI}}$ is 
constant. There $\mu^2$ would be large
enough to allow for perturbative scaling behaviour and the 
multiplication with perturbative Wilson coefficients would make
sense. On the other hand, $\mu^2$ should be small enough to
avoid strong cut-off effects. Indeed, as we have seen, the results 
for the scalar density are rather close to this ideal scenario.
For operators with derivatives the situation is however less
favourable.

Looking at the results for $Z_{\mathrm{RGI}}$
one observes that one gets similar values for all operators 
containing the same number $n_D$ of derivatives.  
They increase with $n_D$. Qualitatively, this behaviour
is reproduced by lattice perturbation theory, and tadpole
improvement enhances the effect, though not in quantitative
agreement with the nonperturbative data. 
In order to facilitate the comparison of the different operators
we have chosen the same vertical scale in the plots for all 
operators with the same number of derivatives although in some 
cases this has the consequence that the results for a few low 
values of $\mu^2$ do not appear on the plot.

\begin{figure}
\vspace*{-2.0cm}
\epsfig{file=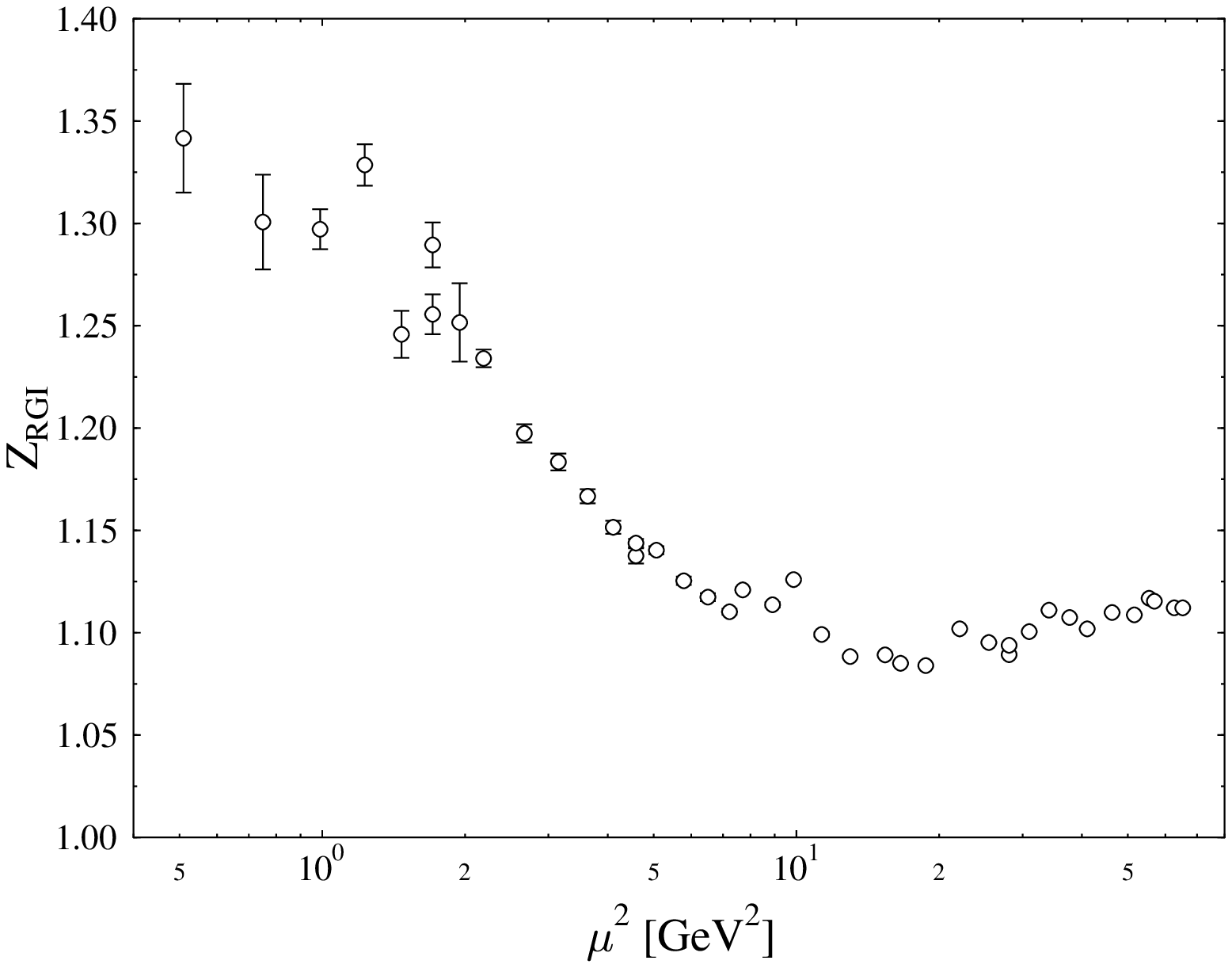,width=13cm} 
\vspace*{-1.0cm}
\caption{$Z_{\mathrm{RGI}}$ for $\cO_{v_{2,a}}$ at $\beta = 6.2$.} 
\label{fig.r.v2a}  
\end{figure}
\begin{figure}
\vspace*{-2.0cm}
\epsfig{file=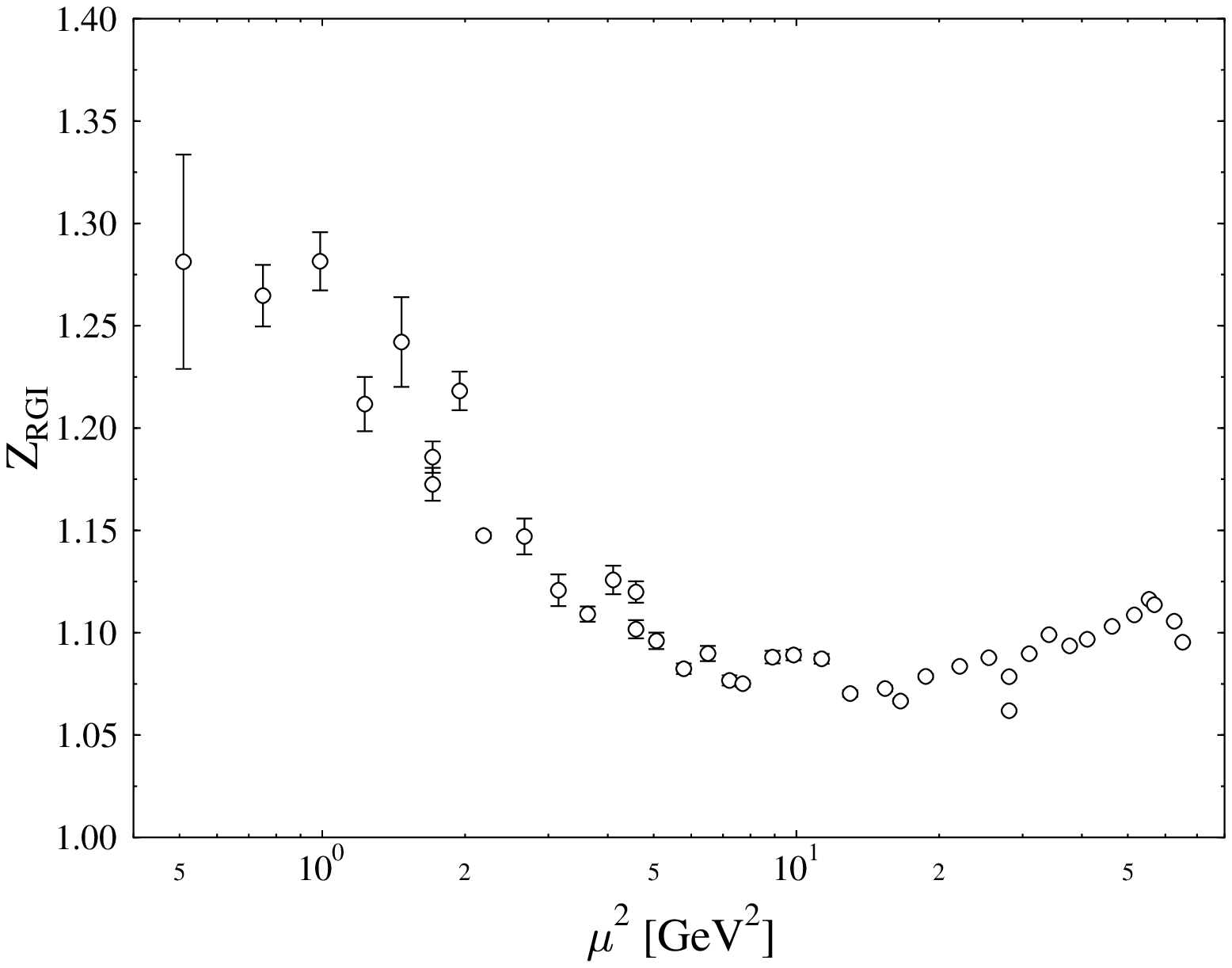,width=13cm} 
\vspace*{-1.0cm}
\caption{$Z_{\mathrm{RGI}}$ for $\cO_{v_{2,b}}$ at $\beta = 6.2$.} 
\label{fig.r.v2b}  
\end{figure}
\begin{figure}
\vspace*{-1.0cm}
\epsfig{file=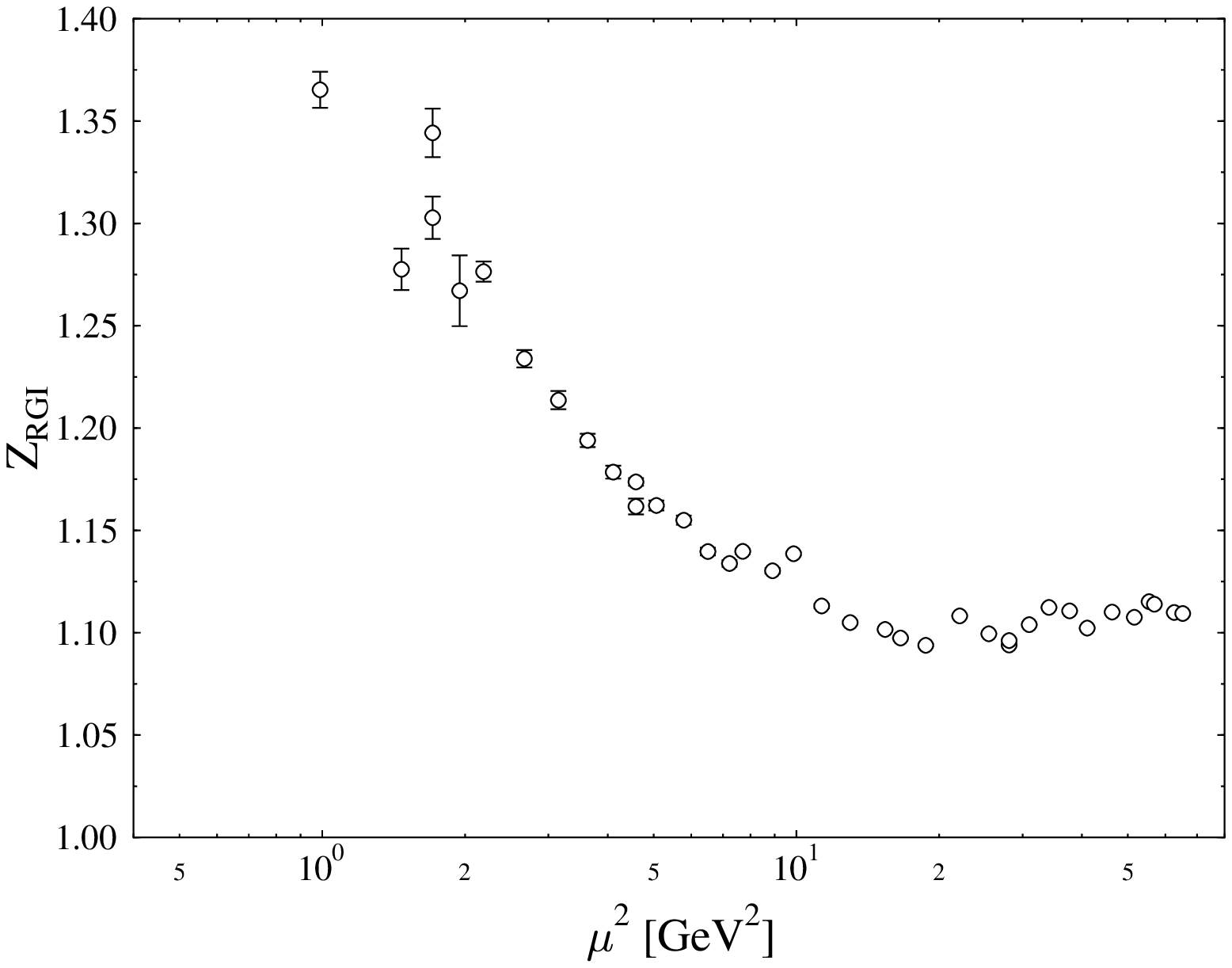,width=13cm} 
\vspace*{-1.0cm}
\caption{$Z_{\mathrm{RGI}}$ for $\cO_{r_{2,a}}$ at $\beta = 6.2$.} 
\label{fig.r.r2a}  
\end{figure}
\begin{figure}
\vspace*{-2.0cm}
\epsfig{file=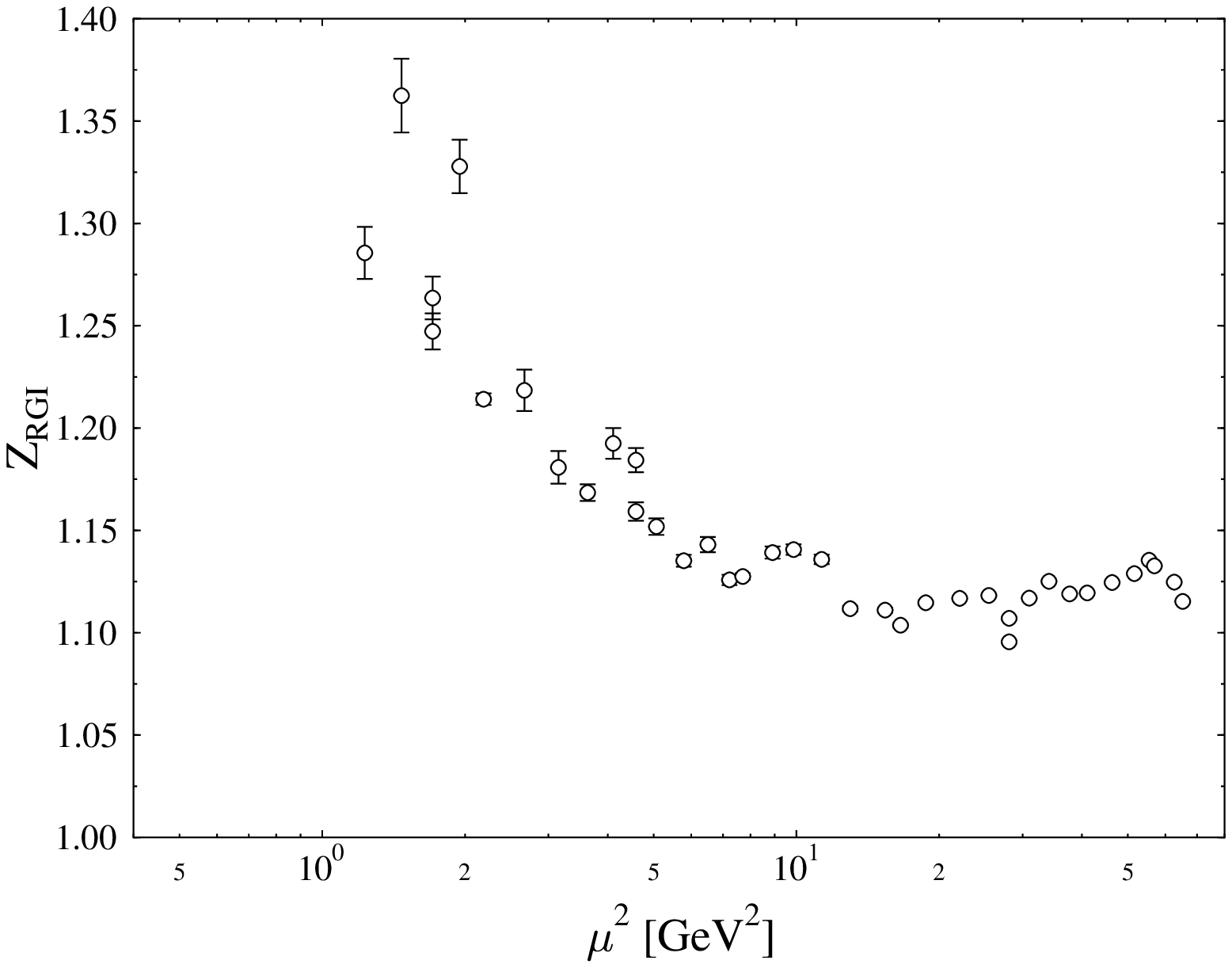,width=13cm} 
\vspace*{-1.0cm}
\caption{$Z_{\mathrm{RGI}}$ for $\cO_{r_{2,b}}$ at $\beta = 6.2$.} 
\label{fig.r.r2b}  
\end{figure}

In Figs.~\ref{fig.r.v2a}-\ref{fig.r.r2b} we plot $Z_{\mathrm{RGI}}$
versus the renormalisation
scale $\mu^2$ for various operators containing one derivative. 
In most cases, a ``flat'' region starts only at 
$\mu^2 \approx 10$ GeV$^2$, where it is hard to believe that
cut-off effects are negligible. Only for $\cO_{v_{2,b}}$ 
(may be also for $\cO_{r_{2,b}}$) the
window might extend to lower $\mu^2$. It should also be noted
that for larger values of $\mu^2$ the scale dependence of $Z$ 
itself (and of $R$) is already rather weak so that approximate
independence of $\mu^2$ is relatively easy to get.  
Let us now compare operators belonging to the same O(4) 
representation in the continuum but transforming differently
under H(4) on the lattice so that their lattice $Z$'s may be 
different although their continuum $Z$'s must coincide.
Remember that we have two pairs
of that kind: $\cO_{v_{2,a}}$, $\cO_{v_{2,b}}$ and 
$\cO_{r_{2,a}}$, $\cO_{r_{2,b}}$.
The results for $\cO_{v_{2,a}}$ lie consistently above those
for $\cO_{v_{2,b}}$, whereas the separation between
$\cO_{r_{2,a}}$ and $\cO_{r_{2,b}}$ is less clear. Having an
anticipatory look at Table~\ref{tab.zmsbar} below we
note that one-loop lattice perturbation theory 
predicts the observed ordering in the case of $\cO_{v_{2,a}}$,
$\cO_{v_{2,b}}$ although it underestimates the size of the difference.
For $\cO_{r_{2,a}}$, $\cO_{r_{2,b}}$, on the other hand, it predicts
a splitting which is roughly one order of magnitude smaller than in
the previous case.  

\begin{figure}
\vspace*{-1.0cm}
\epsfig{file=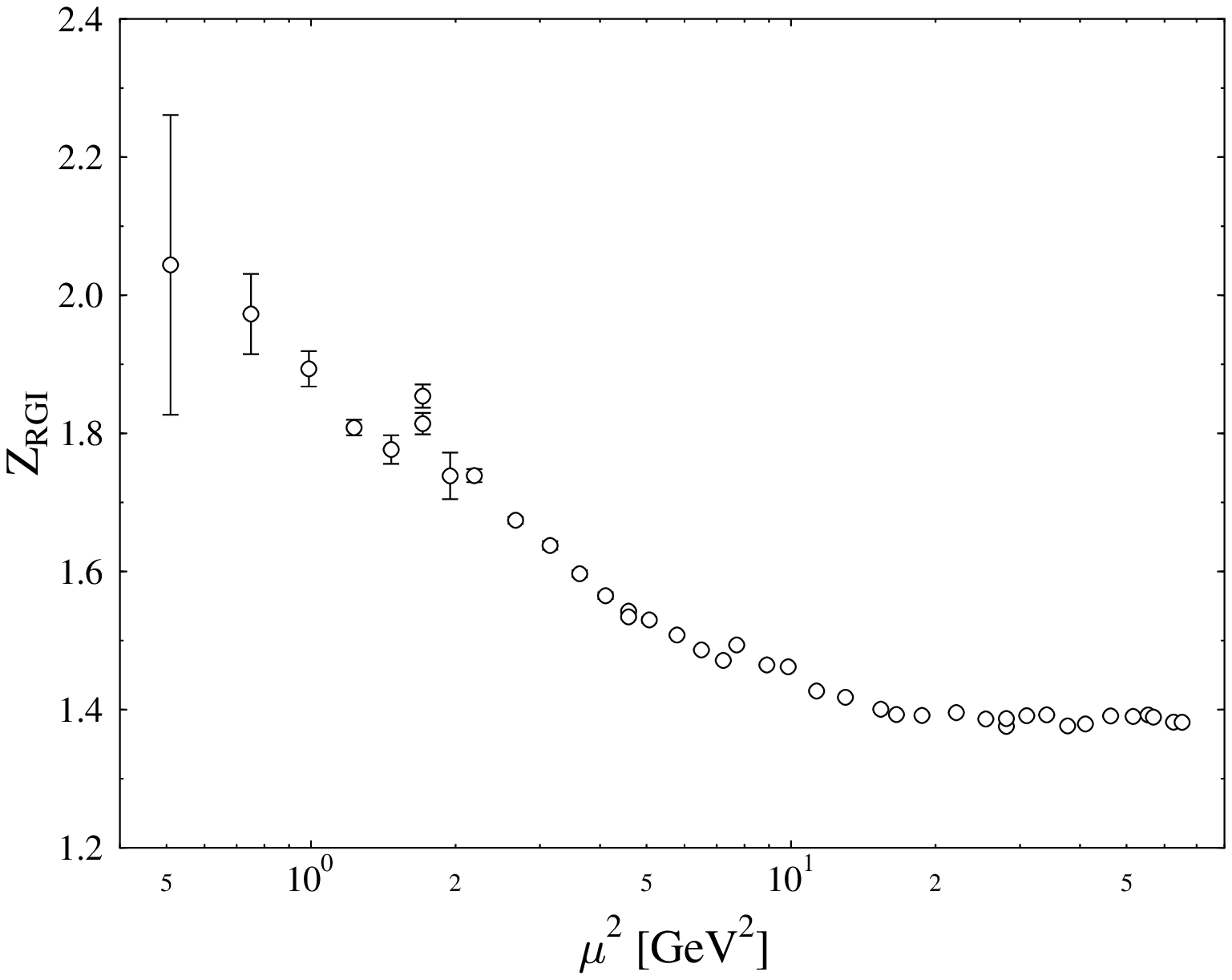,width=13cm} 
\vspace*{-1.0cm}
\caption{$Z_{\mathrm{RGI}}$ for $\cO_{a_2}$ at $\beta = 6.2$.} 
\label{fig.r.a2}  
\end{figure}
\begin{figure}
\vspace*{-2.0cm}
\epsfig{file=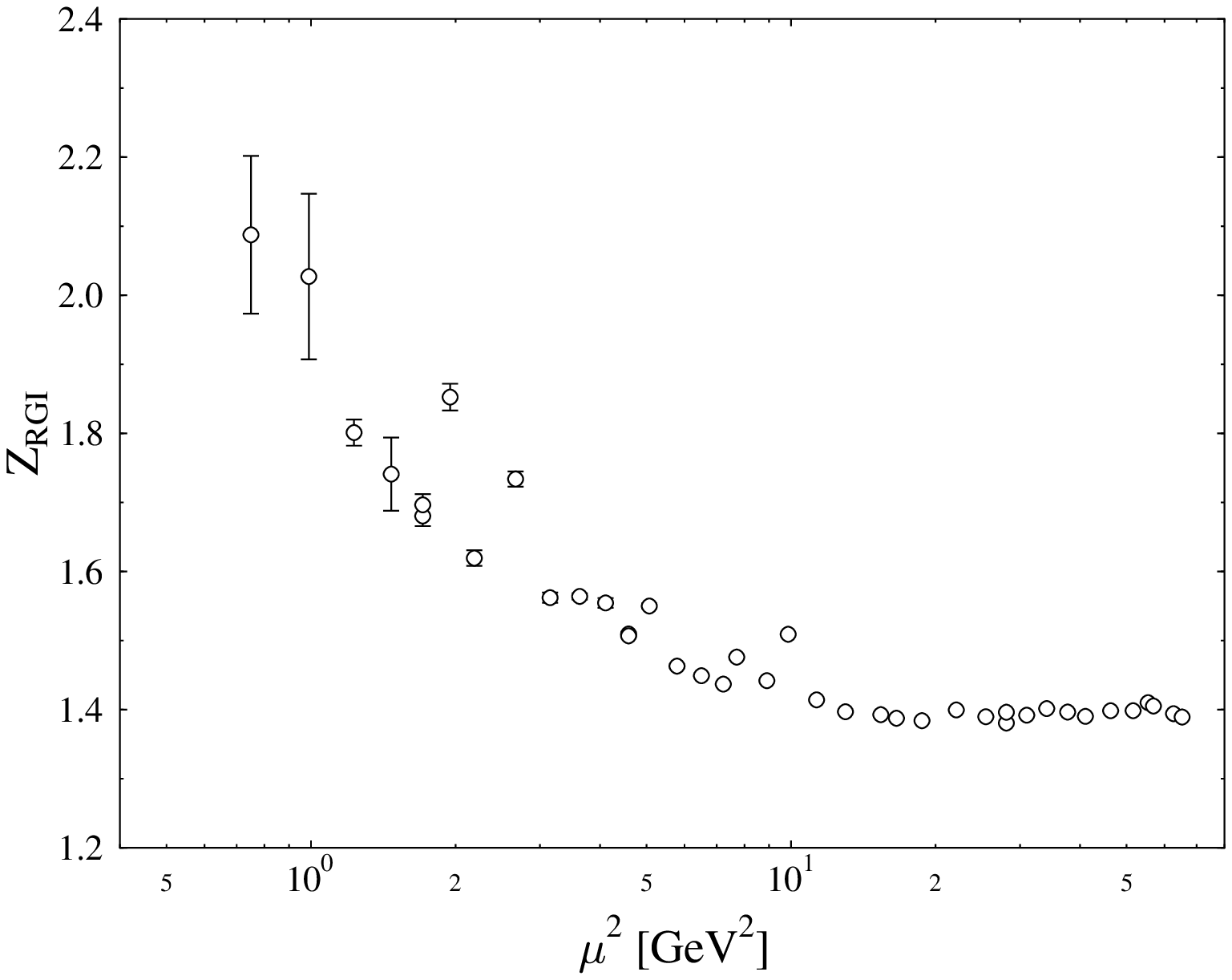,width=13cm} 
\vspace*{-1.0cm}
\caption{$Z_{\mathrm{RGI}}$ for $\cO_{r_3}$ at $\beta = 6.2$.} 
\label{fig.r.r3}  
\end{figure}
\begin{figure}
\vspace*{-1.0cm}
\epsfig{file=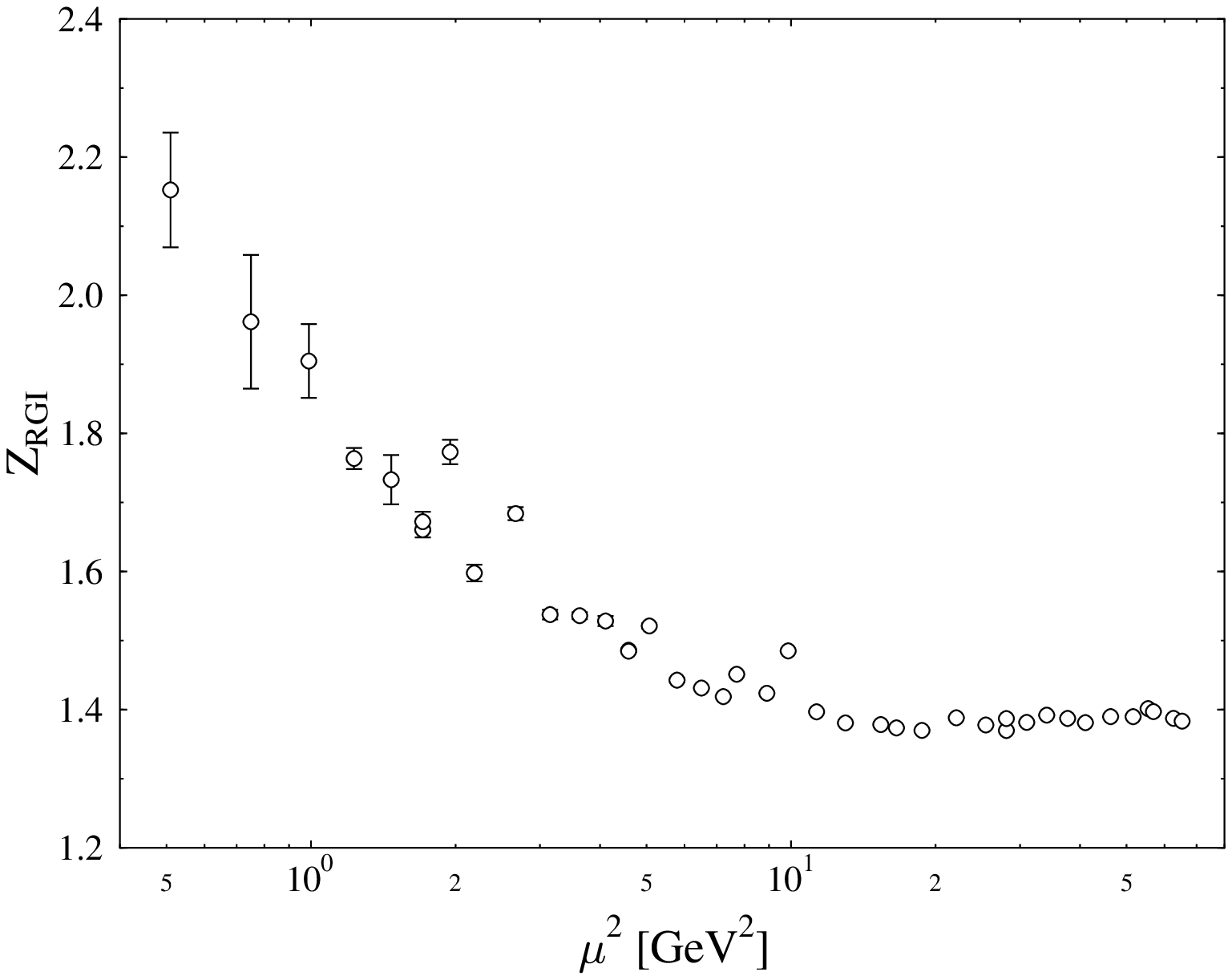,width=13cm} 
\vspace*{-1.0cm}
\caption{$Z_{\mathrm{RGI}}$ for $\cO_{v_3}$ at $\beta = 6.2$.} 
\label{fig.r.v3}  
\end{figure}

In Figs.~\ref{fig.r.a2}-\ref{fig.r.v3} we plot $Z_{\mathrm{RGI}}$
versus the renormalisation
scale $\mu^2$ for three operators containing two derivatives. 
The size of the numbers as well as their spread is larger than in 
the case of the operators with only one derivative. Again, a flat
region is observed only for rather large values of $\mu^2$.
One should also keep in mind that potential mixing problems in the case
of $\cO_{r_3}$ and $\cO_{v_3}$ have been neglected.

\begin{figure}
\vspace*{-2.0cm}
\epsfig{file=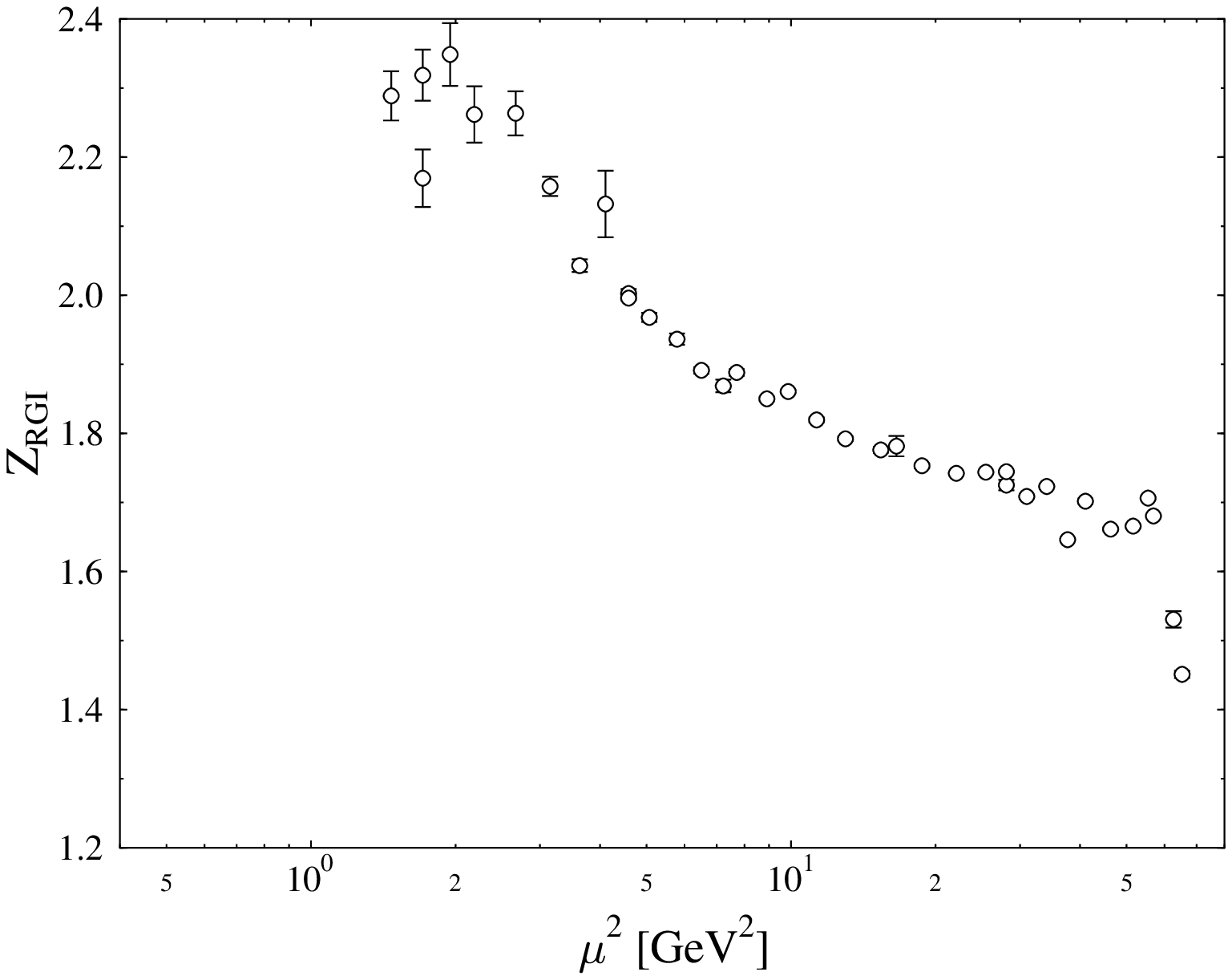,width=13cm} 
\vspace*{-1.0cm}
\caption{$Z_{\mathrm{RGI}}$ for $\cO_{v_4}$ at $\beta = 6.2$.} 
\label{fig.r.v4}  
\end{figure}

In the case of $\cO_{v_4}$, an operator with three derivatives, 
it is simply impossible to identify 
a scaling window as Fig.~\ref{fig.r.v4} shows.
This could be due to the neglected mixing problems
(cf.\ Section~\ref{sec.operators}). If this is true, the
scaling behaviour should be a sensitive test of any attempt to
take the mixing into account. But one would expect anyhow that
cut-off effects might be relatively strong because of the  
large extent of the lattice operator.

\subsection{Comparing $\beta = 6.0$ and $\beta = 6.2$}

\begin{figure}
\vspace*{-2.0cm}
\epsfig{file=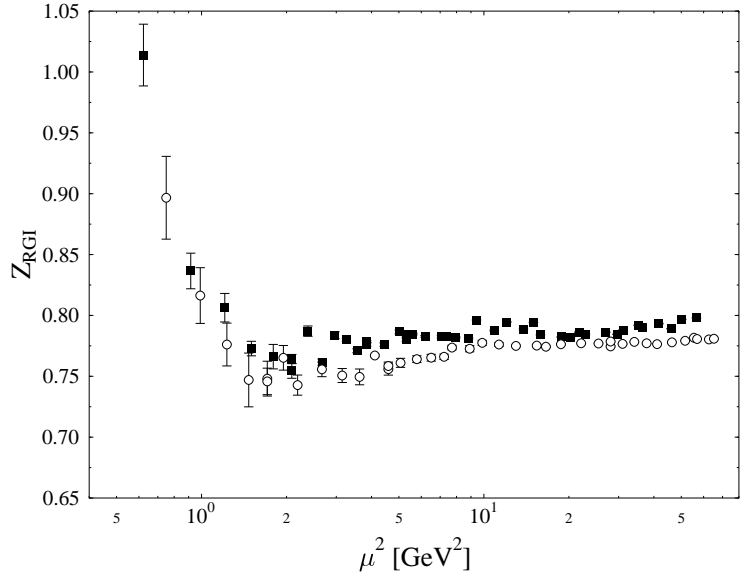,width=13cm} 
\vspace*{-1.0cm}
\caption{$Z_{\mathrm{RGI}}$ for $\cO^S$. The open circles (filled squares) 
         represent the data for $\beta = 6.2$ ($\beta = 6.0$).
         The $\beta = 6.2$ data have been rescaled perturbatively
         to $\beta = 6.0$ by multiplication with (\ref{renfacb}).} 
\label{fig.r2beta.I}  
\end{figure}

\begin{figure}
\vspace*{-1.0cm}
\epsfig{file=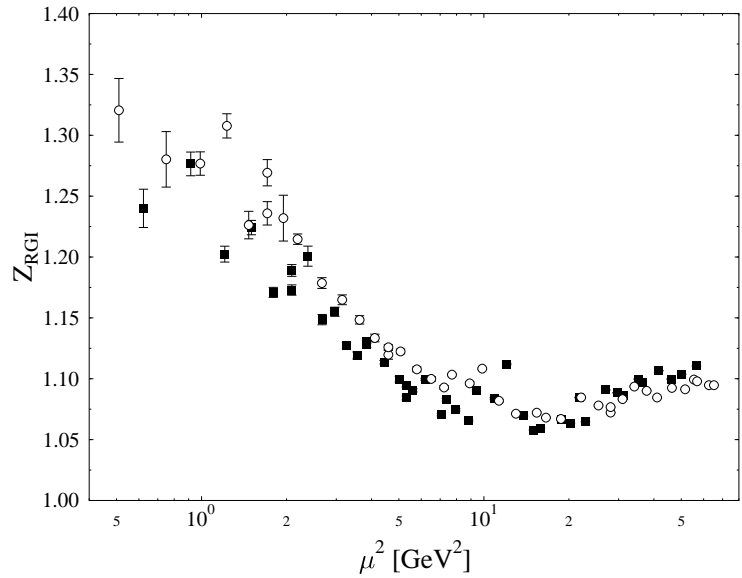,width=13cm} 
\vspace*{-1.0cm}
\caption{The same as Fig.~\ref{fig.r2beta.I}, but for $\cO_{v_{2,a}}$.}
\label{fig.r2beta.v2a}  
\end{figure}

\begin{figure}
\vspace*{-2.0cm}
\epsfig{file=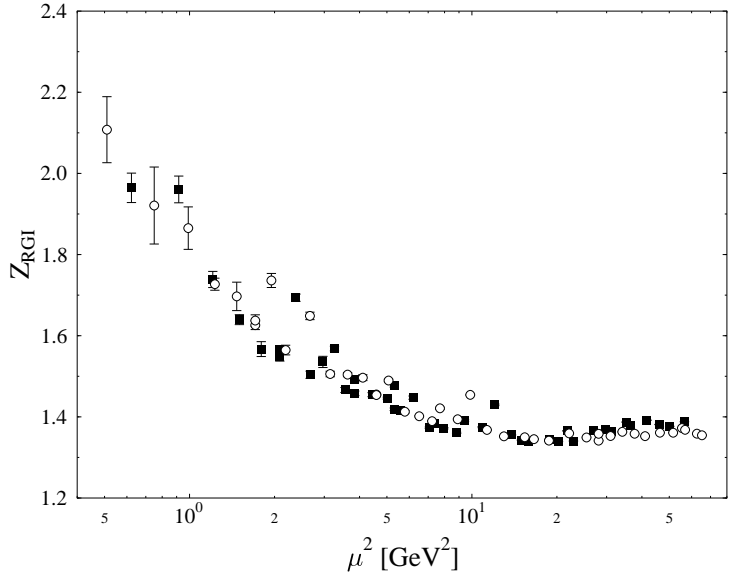,width=13cm} 
\vspace*{-1.0cm}
\caption{The same as Fig.~\ref{fig.r2beta.I}, but for $\cO_{v_3}$.}
\label{fig.r2beta.v3}  
\end{figure}

\begin{figure}
\vspace*{-1.0cm}
\epsfig{file=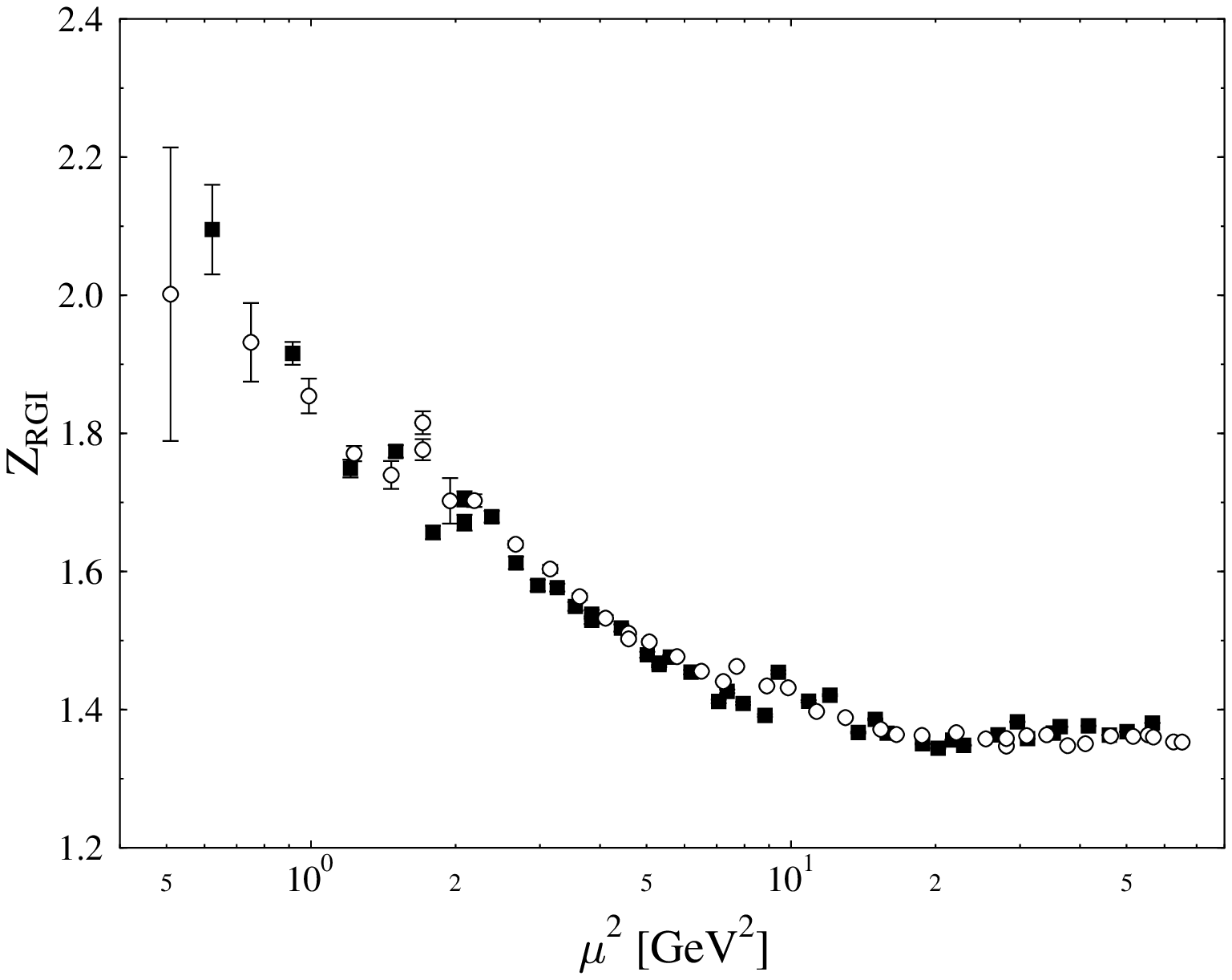,width=13cm} 
\vspace*{-1.0cm}
\caption{The same as Fig.~\ref{fig.r2beta.I}, but for $\cO_{a_2}$.}
\label{fig.r2beta.a2}  
\end{figure}

Let us now compare our results at $\beta = 6.0$ and $\beta = 6.2$
for a few representative operators.
According to the perturbative renormalisation group, the ratio 
of the $Z$'s is given by the $\mu^2$ independent factor 
(\ref{renfacb}) in terms of the bare coupling constants
$g$ and $g^\prime$. 
In Figs.~\ref{fig.r2beta.I}-\ref{fig.r2beta.a2} we plot 
$Z_{\mathrm{RGI}}$ for $\beta = 6.0$ and for $\beta = 6.2$ versus 
the renormalisation scale $\mu^2$. The corresponding plots for
the local vector and axial vector currents have already been 
shown in Figs.~\ref{fig.z2beta.vlh} and \ref{fig.z2beta.vah}.
The $\beta = 6.2$ data have 
been rescaled by multiplication with (\ref{renfacb}), which indeed
moves them closer to the $\beta = 6.0$ results. 
Even outside the scaling windows, the $\beta$ dependence of our
data seems to be compatible with the perturbative expectation
(\ref{renfacb}), at least for most of our operators.
At first sight, this is somewhat puzzling: The perturbative
renormalisation group seems to describe the dependence on the
bare coupling better than the dependence on the momentum scale $\mu$.
However, the ratio 
$a^2(\beta = 6.0)/a^2(\beta = 6.2) \approx 1.84 = 1.36^2$ is not
terribly large and the factor (\ref{renfacb}) differs from one
by at most 2 - 3 \%. Therefore this test of the $\beta$ dependence is
not too stringent.  


Still, it is remarkable that the $\beta$ dependence is close to perturbative
even for rather small values of $\mu^2$. This may be taken as an 
indication that the observed $\mu^2$
dependence is a real physical effect even in regions where it
does not follow the perturbative renormalisation group.

\begin{table}
\caption{Renormalisation factors in the MOM scheme at selected values of
         the renormalisation scale $\mu^2 a^2$ in lattice units.   
         The errors are purely statistical.}
\label{tab.zmom.AVPS}
\vspace{1.0cm}
\begin{center}
\begin{tabular}{cc@{\hspace{1.5cm}}llll}
\hline
$\beta$ & $\mu^2 a^2$ & \multicolumn{1}{c}{$\cO^A_\mu$}  
      & \multicolumn{1}{c}{$\cO^V_\mu$} & \multicolumn{1}{c}{$\cO^P$} 
      & \multicolumn{1}{c}{$\cO^S$} \\
\hline
 {} &     0.318 &     0.809(2)   &     0.645(2)  
                &     0.229(3)   &     0.602(9)  \\
 {} &     0.626 &     0.779(2)   &    0.6658(6)  
                &     0.360(3)   &     0.658(4)  \\
 {} &     0.935 &    0.7803(9)   &    0.6814(8)  
                &     0.431(2)   &     0.679(1)  \\
 {} &     1.320 &    0.7816(7)   &    0.6893(7)  
                &     0.487(1)   &     0.720(1)  \\
6.0 &     1.860 &    0.7850(4)   &    0.7024(6)  
                &    0.5347(7)   &     0.742(1)  \\
 {} &     2.477 &    0.7878(4)   &    0.7088(5)  
                &    0.5772(5)   &    0.7752(4)  \\
 {} &     3.942 &    0.7995(2)   &    0.7286(3)  
                &    0.6320(2)   &    0.8041(2)  \\
 {} &     5.330 &    0.8094(2)   &    0.7543(3)  
                &    0.6731(2)   &    0.8100(2)  \\
 {} &     7.797 &    0.8268(2)   &    0.7761(3)  
                &    0.7127(2)   &    0.8352(2)  \\
\hline
 {} &     0.313 &    0.7954(8)   &    0.6951(3)  
                &     0.357(4)   &     0.605(7)  \\
 {} &     0.587 &     0.797(1)   &    0.7075(3)  
                &     0.463(3)   &    0.6772(9)  \\
 {} &     0.930 &    0.7976(4)   &    0.7189(3)  
                &     0.524(1)   &     0.709(3)  \\
 {} &     1.272 &    0.7994(4)   &    0.7251(3)  
                &     0.566(1)   &     0.738(2)  \\
6.2 &     1.855 &    0.8022(4)   &    0.7352(2)  
                &    0.6078(9)   &    0.7646(8)  \\
 {} &     2.369 &    0.8053(2)   &    0.7435(1)  
                &    0.6328(4)   &    0.7790(7)  \\
 {} &     4.014 &    0.8161(1)   &    0.7649(1)  
                &    0.6877(3)   &    0.8100(4)  \\
 {} &     5.385 &    0.8254(1)   &   0.77760(9)  
                &    0.7150(1)   &    0.8289(2)  \\
 {} &     7.921 &   0.84238(8)   &   0.80054(5)  
                &   0.75556(5)   &    0.8542(2)  \\
\hline
\end{tabular}
\end{center}
\vspace{1.0cm}
\end{table}

\begin{table}
\caption{Renormalisation factors in the MOM scheme at selected values of
         the renormalisation scale $\mu^2 a^2$ in lattice units.   
         The errors are purely statistical.}
\label{tab.zmom.v2v3v4}
\vspace{1.0cm}
\begin{center}
\begin{tabular}{cc@{\hspace{1.5cm}}llll}
\hline
$\beta$ & $\mu^2 a^2$ & \multicolumn{1}{c}{$\cO_{v_{2,a}}$}  
      & \multicolumn{1}{c}{$\cO_{v_{2,b}}$} & \multicolumn{1}{c}{$\cO_{v_3}$} 
      & \multicolumn{1}{c}{$\cO_{v_4}$} \\
\hline
 {} &     0.318 &     1.467(8)   &     1.430(9)  
                &      2.42(3)   &     5.7(1.3)  \\
 {} &     0.626 &     1.352(9)   &     1.294(9)  
                &      2.08(1)   &      2.94(2)  \\
 {} &     0.935 &     1.219(3)   &     1.160(3)  
                &     1.692(7)   &      2.44(1)  \\
 {} &     1.320 &     1.162(2)   &     1.139(3)  
                &     1.596(3)   &      2.31(2)  \\
6.0 &     1.860 &    1.1042(9)   &     1.069(1)  
                &     1.454(2)   &     2.000(4)  \\
 {} &     2.477 &    1.0999(7)   &    1.0486(5)  
                &     1.431(2)   &     1.879(8)  \\
 {} &     3.942 &    1.0366(4)   &    1.0527(8)  
                &    1.3167(9)   &     1.701(3)  \\
 {} &     5.330 &    1.0254(2)   &    1.0094(2)  
                &    1.2753(5)   &      1.64(1)  \\
 {} &     7.797 &    1.0271(2)   &    1.0107(2)  
                &    1.2615(4)   &     0.909(2)  \\
\hline
 {} &     0.313 &     1.408(5)   &     1.304(2)  
                &      1.98(2)   &      3.02(5)  \\
 {} &     0.587 &     1.236(3)   &     1.210(7)  
                &     1.731(8)   &      2.53(6)  \\
 {} &     0.930 &     1.158(2)   &     1.128(4)  
                &     1.529(2)   &     2.084(5)  \\
 {} &     1.272 &     1.128(1)   &     1.102(3)  
                &    1.4696(6)   &     1.950(3)  \\
6.2 &     1.855 &    1.0774(7)   &     1.057(2)  
                &    1.3709(5)   &     1.799(3)  \\
 {} &     2.369 &    1.0583(5)   &     1.038(1)  
                &    1.3319(4)   &      1.73(1)  \\
 {} &     4.014 &    1.0313(2)   &    1.0190(7)  
                &    1.2680(4)   &     1.583(7)  \\
 {} &     5.385 &    1.0321(2)   &    1.0181(5)  
                &    1.2543(3)   &     1.466(2)  \\
 {} &     7.921 &    1.0210(2)   &    1.0211(4)  
                &    1.2296(5)   &     1.463(1)  \\
\hline
\end{tabular}
\end{center}
\vspace{1.0cm}
\end{table}

\begin{table}
\caption{Renormalisation factors in the MOM scheme at selected values of
         the renormalisation scale $\mu^2 a^2$ in lattice units.   
         The errors are purely statistical.}
\label{tab.zmom.a2r2r3}
\vspace{1.0cm}
\begin{center}
\begin{tabular}{cc@{\hspace{1.5cm}}llll}
\hline
$\beta$ & $\mu^2 a^2$ & \multicolumn{1}{c}{$\cO_{a_2}$}  
      & \multicolumn{1}{c}{$\cO_{r_{2,a}}$} 
      & \multicolumn{1}{c}{$\cO_{r_{2,b}}$} 
      & \multicolumn{1}{c}{$\cO_{r_3}$} \\
\hline
 {} &     0.318 &      2.44(2)   &     1.539(5)  
                &      1.60(1)   &      2.52(3)  \\
 {} &     0.626 &      2.06(1)   &     1.382(7)  
                &     1.417(10)  &      2.19(1)  \\
 {} &     0.935 &     1.796(7)   &     1.259(4)  
                &     1.235(3)   &     1.724(7)  \\
 {} &     1.320 &     1.636(5)   &     1.189(3)  
                &     1.211(3)   &     1.629(4)  \\
6.0 &     1.860 &     1.503(3)   &     1.129(1)  
                &    1.1230(9)   &     1.478(2)  \\
 {} &     2.477 &     1.494(3)   &    1.1162(7)  
                &    1.1040(5)   &     1.455(2)  \\
 {} &     3.942 &     1.360(1)   &    1.0458(4)  
                &    1.0942(8)   &     1.331(1)  \\
 {} &     5.330 &    1.2852(6)   &    1.0282(2)  
                &    1.0387(3)   &    1.2869(6)  \\
 {} &     7.797 &    1.2773(6)   &    1.0308(2)  
                &    1.0363(2)   &    1.2710(5)  \\
\hline
 {} &     0.313 &      2.17(1)   &     1.457(6)  
                &     1.380(3)   &      2.01(1)  \\
 {} &     0.587 &     1.776(5)   &     1.265(3)  
                &     1.281(8)   &     1.760(8)  \\
 {} &     0.930 &     1.594(2)   &     1.181(2)  
                &     1.183(4)   &     1.548(2)  \\
 {} &     1.272 &    1.5152(8)   &     1.145(1)  
                &     1.154(3)   &    1.4884(8)  \\
6.2 &     1.855 &    1.4124(5)   &    1.0938(9)  
                &     1.098(1)   &    1.3869(5)  \\
 {} &     2.369 &    1.3564(3)   &    1.0704(6)  
                &    1.0737(10)  &    1.3455(4)  \\
 {} &     4.014 &    1.2782(3)   &    1.0358(3)  
                &    1.0460(6)   &    1.2780(3)  \\
 {} &     5.385 &    1.2479(3)   &    1.0351(2)  
                &    1.0417(4)   &    1.2627(3)  \\
 {} &     7.921 &    1.2241(3)   &    1.0194(2)  
                &    1.0386(3)   &    1.2372(4)  \\
\hline
\end{tabular}
\end{center}
\vspace{1.0cm}
\end{table}

\begin{table}
\caption{Renormalisation factors in the $\overline{\mathrm{MS}}$ scheme
         for $\beta = 6.0$ at $\mu^2 = a^{-2} = 3.8$~GeV$^2$ and
         for $\beta = 6.2$ at $\mu^2 = a^{-2} = 7$~GeV$^2$ .The errors
         of the nonperturbative values $Z_\cO$ are purely statistical.} 
\label{tab.zmsbar}
\vspace{1.0cm}
\begin{center}
\begin{tabular}{crlccrlcc}
\hline
$\cO$ & \hspace{0.5cm} & \multicolumn{1}{c}{$Z_\cO$} 
      & $Z^{\mathrm{ti}}_\cO$ & $Z^{\mathrm{pert}}_\cO$ 
      & \hspace{0.5cm} & \multicolumn{1}{c}{$Z_\cO$} 
      & $Z^{\mathrm{ti}}_\cO$ & $Z^{\mathrm{pert}}_\cO$ \\
\hline
 {}   & {}  & \multicolumn{3}{c}{$\beta = 6.0$} & {}
             & \multicolumn{3}{c}{$\beta = 6.2$} \\
\hline
$\cO_{v_{2,a}}$ & {} &    1.1332(29)  &    0.9731  &    0.9892 & {}
                     &    1.0732(19)  &    0.9759  &    0.9895  \\
$\cO_{v_{2,b}}$ & {} &    1.0836(24)  &    0.9462  &    0.9784 & {}
                     &    1.0426(36)  &    0.9518  &    0.9791  \\
$\cO_{v_3}$     & {} &    1.4973(67)  &    1.1933  &    1.1024 & {}
                     &    1.3458(15)  &    1.1778  &    1.0991  \\
$\cO_{v_4}$     & {} &    2.0042(121) &    1.5021  &    1.2299 & {}
                     &    1.7497(85)  &    1.4565  &    1.2225  \\
$\cO_{a_2}$     & {} &    1.5409(64)  &    1.1930  &    1.1023 & {}
                     &    1.3962(16)  &    1.1776  &    1.0990  \\
$\cO_{r_{2,a}}$ & {} &    1.1615(33)  &    0.9927  &    0.9971 & {}
                     &    1.0956(18)  &    0.9935  &    0.9972  \\
$\cO_{r_{2,b}}$ & {} &    1.1541(24)  &    0.9965  &    0.9986 & {}
                     &    1.0912(36)  &    0.9968  &    0.9986  \\
$\cO_{r_3}$     & {} &    1.5292(68)  &    1.2108  &    1.1086 & {}
                     &    1.3628(15)  &    1.1934  &    1.1051  \\
$\cO^S$         & {} &    0.7718(16)  &    0.8209  &    0.8906 & {}
                     &    0.7897(29)  &    0.8337  &    0.8942  \\
$\cO^P$         & {} &    0.4934(24)  &    0.6430  &    0.8092 & {}
                     &    0.5888(15)  &    0.6731  &    0.8154  \\
$\cO^V_\mu$     & {} &    0.6833(8)   &    0.6795  &    0.8259 & {}
                     &    0.7206(3)   &    0.7060  &    0.8315  \\
$\cO^A_\mu$     & {} &    0.7821(9)   &    0.7684  &    0.8666 & {}
                     &    0.7978(4)   &    0.7864  &    0.8709  \\
\hline
\end{tabular}
\end{center}
\vspace{1.0cm}
\end{table}

\subsection{Compilation of results} 

In Tables~\ref{tab.zmom.AVPS} - \ref{tab.zmom.a2r2r3} we present our
results for $Z_{\cO}$ in the MOM scheme at selected values of the 
renormalisation scale. It is to be expected that the systematic
uncertainties (in particular cut-off effects) are considerably
larger than the quoted statistical errors. A reasonable measure
for the size of the cut-off effects might be the width of the band
of data points which is typically of the order of a few percent for not
too small values of $\mu^2$. 

In Table~\ref{tab.zmsbar} we give the nonperturbative values for
$Z_\cO$ in the $\overline{\mathrm{MS}}$ scheme together with
$Z^{\mathrm{ti}}_\cO$ and $Z^{\mathrm{pert}}_\cO$ for
$\mu^2 = a^{-2}$. In order to determine the nonperturbative $Z_\cO$'s
for this value of $\mu^2$ we have interpolated linearly in $\ln \mu^2$
between the two neighbouring results. The given error is the larger
of the two statistical errors. Disregarding the results at other 
values of $\mu^2$ we see that, with the exception of the operators
containing one covariant derivative, tadpole improvement works
in the right direction. In the case of the local vector and axial
vector currents it even overshoots somewhat. However, only in these
latter cases the improvement is quantitatively satisfactory.

\begin{figure}
\vspace*{-2.0cm}
\epsfig{file=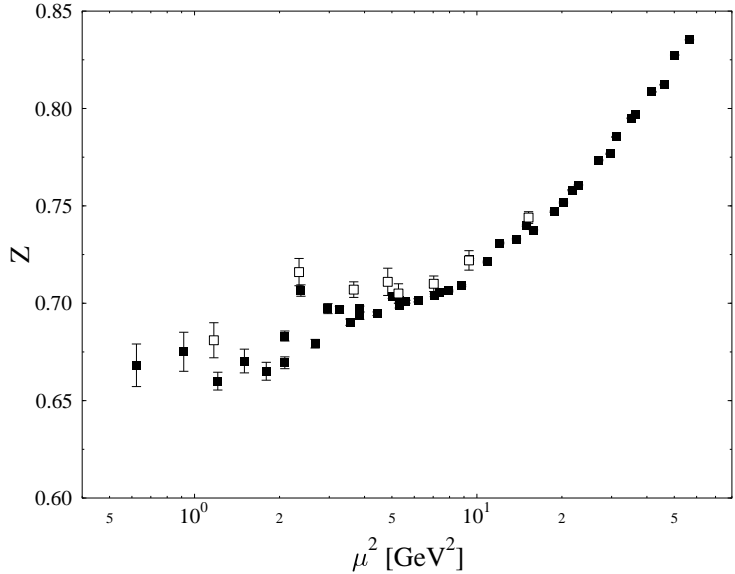,width=13cm} 
\vspace*{-1.0cm}
\caption{$Z$ for the local vector current at $\beta = 6.0$.
         The filled squares represent our data averaged
         over the four components of the current. The open
         squares are results taken from Ref.~\cite{gimenez}.}      
\label{fig.zgim.vln}  
\end{figure}

The renormalisation factors for the operators without derivatives
($\cO^S $, $\cO^P $, $\cO^V_\mu $, $\cO^A_\mu $) have been calculated
with similar nonperturbative methods by other groups, too. 
However, most of these studies
use either an improved fermionic action or work at different values
of $\beta$ so that a direct comparison is impossible. We are aware
of only one other paper dealing with Wilson fermions at $\beta = 6.0$
and $\beta = 6.2$ \cite{gimenez}. Let us compare our results as
given in Table~\ref{tab.zmom.AVPS} with Table~3 of Ref.~\cite{gimenez}.
\begin{itemize}
\item In the case of the local vector and axial vector currents
      we find some discrepancies, which, however, 
      decrease with increasing $\mu^2$. At $\beta = 6.2$ we obtain
      consistent results for the larger values of $\mu^2$. 
      The deviations could be due to the fact that
      we use only the transverse components of the 
      currents. Indeed, averaging over the four $Z$'s 
      determined from the four components of the local vector 
      current by the method of Section~\ref{sec.method}, we 
      arrive at Fig.~\ref{fig.zgim.vln},
      where also the results of Ref.~\cite{gimenez} are shown.
      We observe satisfactory agreement as well as a relatively
      large spread of the data, which is not present in our method
      based on the transverse components (see Fig.~\ref{fig.z2beta.vlh}).
\item The pseudoscalar density shows a different behaviour.
      Our $Z$'s are consistently smaller for $\beta = 6.0$, 
      but always larger for $\beta = 6.2$ (although the differences
      are not large). Here the chiral extrapolation might be
      responsible for the discrepancies. Especially for the lower
      values of $\mu^2 a^2$ we observe deviations from a linear
      $1/\kappa$ dependence, which make the extrapolation somewhat
      questionable and ambiguous. 
\item For the scalar density we find satisfactory agreement if 
      $\mu^2 a^2$ is not too small. 
      Like in our data, the ratio $Z_{\cO^P}/Z_{\cO^S}$ does not
      become constant even for the largest values of $\mu^2$
      considered in \cite{gimenez}.
\end{itemize}
In all cases one should keep in mind that the lattice used in
\cite{gimenez} at $\beta = 6.2$ is smaller than ours. Furthermore,
the momenta may differ in direction even if they are close in
length, hence part of the differences could also be lattice
artifacts.

The renormalisation constant of the local vector current can 
also be computed from its correlation functions with hadron sources by
comparing with the corresponding correlation functions of the conserved
vector current, whose renormalisation constant is known to
be one. A collection of results obtained with various versions of 
this method at $\beta = 6.0$ is given by Sachrajda \cite{sach} 
(cf.\ also \cite{maimart}), albeit without extrapolation 
to the chiral limit. Since the quark mass dependence is rather
mild, we may nevertheless compare with our numbers as displayed
in Fig.~\ref{fig.z2beta.vlh}. The values
given in Ref.~\cite{sach} cluster around 0.73 (computed from
three-point functions) and 0.57 (computed from two-point functions).
Sachrajda attributes this large difference to discretization effects.
This is in accordance with our interpretation of the strong scale
dependence of the $Z$ for the local vector current 
(see Subsection~\ref{subsec.veccur}). More recently, the JLQCD
collaboration has applied this method with Wilson fermions at
$\beta=5.9$, 6.1, and 6.3 \cite{jlqcd}. Interpolating their results
\cite{buckow} one finds 0.57 at $\beta=6.0$ and 0.62 at $\beta=6.2$.  

The renormalisation of the local vector current and the axial vector
current can also be studied by means of the Ward identity method 
\cite{bochi}. However, the values obtained for Wilson fermions 
at $\beta = 6.0$ by Maiani and Martinelli \cite{maimart} (0.79(4) for the
vector current and 0.85(7) for the axial vector current)
refer to point-split lattice currents and are therefore not
immediately comparable with our numbers. On the other hand, for the 
ratio $Z_{\cO^V_\mu}/Z_{\cO^A_\mu}$ of the renormalisation constants
of the local currents they give the values 0.85(5)
and 0.79(5) which are lower than (though not completely incompatible 
with) our result of 0.87 obtained at $\mu^2 a^2 = 1$.  
From the $Z$'s which the JLQCD collaboration has calculated 
for the axial vector current by the Ward identity method
at $\beta=5.9$, 6.1, and 6.3 one gets by interpolation \cite{buckow} 
approximately 0.75 ($\beta=6.0$) and 0.77 ($\beta=6.2$), which
is somewhat lower than our values for $\mu^2 a^2 = 1$.

There have been various proposals to increase the accuracy in
the determination of renormalisation constants, in particular
by using modified versions of tadpole improvement. For 
$Z_{\cO^V_\mu}$, $Z_{\cO^A_\mu}$, and the ratio $Z_{\cO^P}/Z_{\cO^S}$
these are discussed in Ref.~\cite{crisafulli}, where also a
comparison with results from the Ward identity method is given. 

\section{Conclusions}

We have presented a comprehensive study of nonperturbative renormalisation
for various operators in the framework of lattice QCD.
In particular, twist-2 operators appearing in unpolarised
(polarised) deep-inelastic scattering 
have been studied for all spins $\leq 4$ ($\leq 3$).
We worked with standard Wilson fermions at two different
lattice spacings. Due to the use of momentum sources we achieved a 
high statistical accuracy so that systematic uncertainties
(cut-off effects, in particular) are clearly
visible. In order to make contact with perturbation theory one would
like to identify scaling windows where the renormalisation scale $\mu$ is
large enough to make perturbation theory trustworthy but small enough
to allow to neglect cut-off effects. In this respect some operators
did not follow our (naive) expectations. Whereas the axial vector current
and the scalar density behave as expected, the local vector
current and the pseudoscalar density do not show perturbative scaling.
A second thought, however, revealed that there might be good reasons
for this behaviour related to the spontaneous breakdown of chiral
symmetry and there is nothing fundamentally wrong.

On the other hand, the operators containing one or more covariant
derivatives seem to approach perturbative scaling only for rather
large values of the renormalisation scale where cut-off effects
might influence the results. Can we understand these deviations 
from perturbative scaling which seem to persist up to 
$\mu^2 \approx 10$GeV$^2$? If the observed scaling with $\beta$
is not only an artifact of our two $\beta$ values being
relatively close to each other, the deviations from scaling in $\mu$
should be interpreted as real physics, though in a finite volume.
(Recall that our lattices at the two $\beta$ values have roughly 
the same physical size.) So one possibility would be finite-size 
effects, although the experience with other observables
suggests that they should be small.
But genuinely nonperturbative effects cannot be ruled out,
and one might also think of renormalons. In any case, one does
not feel very comfortable when combining these nonperturbative
$Z$'s with perturbative Wilson coefficients using a scale of,
say, $\mu^2 = 4$GeV$^2$. 

One remedy would be a nonperturbative calculation of the 
Wilson coefficients. 
But one can also try to reach larger values of $\mu^2$. In order
to avoid strong cut-off effects one would then have to work at
smaller lattice spacings, i.e.\ at $\beta > 6.2$. However,
this raises a new problem. At present, we have been careful to  
work with lattice sizes $L$ large enough that we believe 
finite-size effects are not serious. But if we try to reach much
smaller $a$'s it soon becomes impossibly expensive to keep the
physical lattice size sufficiently large. So we need a reliable
method of measuring $Z$ nonperturbatively, even when finite-size
effects are not negligible. Such a scheme has been suggested in
Ref.~\cite{luescher}. The key point is to note that although
the Green functions depend on $L$, the renormalisation constants
do not. This means that we can nonperturbatively measure the 
ratios of $Z$'s at different lattice spacings by looking at
the ratios of the corresponding bare Green functions, both
measured on lattices of the same physical size, and with bare
masses chosen so that the renormalised mass is the same in both 
cases. Note that to apply this scheme, we need good knowledge
of the lattice spacing as a function of $\beta$, which may be
a problem. 

Further improvement could come from continuum perturbation theory. 
In the case of the scalar density the use of the two-loop formula for
$Z^{\overline{\mathrm{MS}}}_{\mathrm{MOM}}$ and of the
three-loop anomalous dimension improves the (approximate) scale 
independence of $Z_{\mathrm{RGI}}$. For the operators with derivatives
we only have the two-loop anomalous dimension and a one-loop calculation
of $Z^{\overline{\mathrm{MS}}}_{\mathrm{MOM}}$. Recently, however,
first steps towards a three-loop computation of the anomalous dimension
have been undertaken \cite{neerven}. From the results already obtained,
a two-loop expression for
$Z^{\overline{\mathrm{MS}}}_{\mathrm{MOM}}$ can be extracted,
unfortunately only in the Feynman gauge, whereas we would need the
Landau gauge.

\begin{ack}
This work is supported by the Deutsche Forschungsgemeinschaft and 
by BMBF. The numerical calculations were performed on the Quadrics 
computers at DESY-Zeuthen. We wish to thank the operating staff 
for their support.
\end{ack}


\end{document}